\documentclass[fleqn,usenatbib]{mnras}

\usepackage{newtxtext,newtxmath}
\usepackage[T1]{fontenc}
\usepackage{ae,aecompl}

\usepackage{graphicx}	% Including figure files
\usepackage{amsmath}	% Advanced maths commands
\usepackage{bm}
\newcommand{\appropto}{\mathrel{\vcenter{
  \offinterlineskip\halign{\hfil$##$\cr
    \propto\cr\noalign{\kern2pt}\sim\cr\noalign{\kern-2pt}}}}}

\title[A Geometric Probe of Cosmology -- II.]{A Geometric Probe of Cosmology -- II. Gravitational Lensing Time Delays and Quasar Reverberation Mapping Revisited}

\author[Ng]{
Angela L.H. Ng\thanks{E-mail: angela.ng@sydney.edu.au}
\\
School of Physics, A28, The University of Sydney, NSW 2006, Australia\\\
}

\date{Accepted XXX. Received YYY; in original form ZZZ}

\pubyear{2023}

\begin{document}
\label{firstpage}
\pagerange{\pageref{firstpage}--\pageref{lastpage}}
\maketitle

\begin{abstract}
The time delay between images of strongly gravitationally lensed quasars is an established cosmological probe. Its limitations, however, include uncertainties in the assumed mass distribution of the lens. We re-examine the methodology of a prior work presenting a geometric probe of cosmology independent of the lensing potential which considers differential time delays over images, originating from spatially-separated photometric signals within a strongly lensed quasar. We give an analytic description of the effect of the differential lensing on the emission line spectral flux for axisymmetric Broad Line Region geometries, with the inclined ring or disk, spherical shell, and double cone as examples. The proposed method is unable to recover cosmological information as the observed time delay and inferred line-of-sight velocity do not uniquely map to the three-dimensional position within the source.
\end{abstract}


%%%%%%%%%%%%%%%%%%%%%%%%%%%%%%%%%%%%%%%%%%%%%%%%%%

%%%%%%%%%%%%%%%%% BODY OF PAPER %%%%%%%%%%%%%%%%%%

\section{Introduction}

 The time delay due to geometric and gravitational differences in the light paths between multiple images of strongly gravitationally lensed quasars is a direct measurement of cosmological distances, and therefore of cosmological parameters \citep{Refsdal1964, BlandfordNarayan1992}. Time delay cosmography is now an established independent method by which cosmological information can be determined, e.g. \citet{Tewes2013, Courbin2018, Birrer2019a}, and \citet[H0LiCOW;][]{Suyu2017} with a current 2.4 per cent precision measurement of $H_0$ \citep{Wong2020}. However, gravitational lens modelling depends implicitly upon the assumed underlying mass distribution. In particular, the ``mass sheet degeneracy'' involving the degeneracy of ill-constrained lens models leads to systematic uncertainties in estimations of $H_0$ \citep{Falco1985, Saha2000, Kochanek2020, Birrer2020, Chen2020}.

In a first paper \citep{Ng2020} we presented a geometric probe of cosmology independent of the lensing potential, by considering differential time delays over images, originating from spatially-separated photometric signals within a strongly lensed quasar. In the original work, we presented the predicted signal integrated across all wavelengths as a function of time, from considering a special case of a thin face-on ring geometry of the Broad Line Region (BLR) of the quasar. In this paper we re-examine the methodology and demonstrate some significant difficulties, most notably under-determination. We also consider analytically the picture in the line-of-sight velocity (i.e. spectral) and time delay space, referred to as a ``velocity-delay map''; as well as the signals obtained at any spectral or temporal slice for axisymmetric BLR geometries with the inclined ring or disk, spherical shell, and double cone as examples.

The paper is organised as follows: we briefly review the definitions and the background in Section \ref{sec:background} and give an overview of the methodology and difficulties in Section \ref{sec:methodanddifficulties}. We look at the effect of the differential lensing on the velocity-delay maps of different BLR geometries in Section \ref{sec:velocitydelaymaps} and demonstrate the insensitivity of the required BLR parameter estimation to the differential lensing in Section \ref{sec:parameterestimation}.

\section{Background} \label{sec:background}

An observer sees a photometric signal from an unlensed point source at line-of-sight physical distance $\xi_z$ and position $\bm{\beta}$ on the sky, at time $t=t_i$. When the source is lensed, the observer sees this signal in a given lensed image X at a position $\bm{\theta}_X$ on the sky and at time $t = t_i + \tau_X$, where $\tau_X$ is the time delay \citep[e.g.][]{Schneider1992, Blandford1986, BlandfordNarayan1992}:
\begin{equation}
    \tau_X \equiv \tau \left(\bm{\theta}_X , \bm{\beta}, \xi_z \right) = \frac{D}{c} ( 1 + z_d ) \left(\frac{1}{2} ( \bm{\theta}_X - \bm{\beta})^2 - \psi ( \bm{\theta}_X )\right). \label{tauX}
\end{equation}
Here $D= \frac{D_d D_s}{D_{ds}} \propto H_0^{-1}$ is the ``lensing distance'' or more commonly the ``time-delay distance'', a ratio of angular diameter distances (subscript $d$, $s$, $ds$ denoting the angular diameter distance from the observer to the lens, to the source and from the lens to the source, respectively). The lensing distance is thus the factor containing all the cosmological information. We denote the speed of light by $c$, and the redshift of the lensing mass by $z_d$. The two-dimensional vector positions of image X in the lens plane, and the source in the source plane, are given by $\bm{\theta}_X$ and $\bm{\beta}$ respectively; scaled such that their magnitudes are the observed angular positions relative to the observer-lens axis. The dimensionless lensing potential is denoted by $\psi$.

Although the time delay $\tau$ is not an observable quantity, the time delays between images of strongly lensed quasars may be measured and is a conventional method employed to test cosmology. The observed time delay between two images A and B are given by
\begin{align}
\begin{split}
\Delta \tau_{BA} \equiv& \, \tau_B - \tau_A\\
=& \, \frac{D}{c} (1 +z_d) \Big(\frac{1}{2}  \left(\bm{\theta}^2_B - \bm{\theta}^2_A \right) + (\bm{\theta}_A - \bm{\theta}_B) \cdot \bm{\beta} \\
&- \psi (\bm{\theta}_B) + \psi (\bm{\theta}_A)  \Big). \label{tauBA}
\end{split}
\end{align}

Since Equation \eqref{tauBA} is dependent on the  dimensionless lensing potential $\psi $ however, time delay measurements are limited by the assumptions and accuracy of the lens model. Simple gravitational lens systems are rare, and observational constraints on the lens model are limited due to the existence, typically, of only two or four images per system \citep{Schneider2013, Birrer2019a}.

In Paper I \citep{Ng2020}, we instead consider lensed quasars as a finite rather than point source in combination with reverberation mapping \citep{Blandford1982, Peterson1993, Cackett2021} to determine quasar structure. Broad emission lines are a feature of quasar spectra and correspond to gas clouds, known as the Broad Line Region (BLR), surrounding the central emitting accretion disk at some distance. Photons travelling outwards from the central source are absorbed and re-emitted by the BLR gas. The broad emission lines therefore respond to variations in the continuum luminosity of the central source with a time delay determined by the BLR geometry.

For an unlensed quasar, if the continuum emission (across all wavelengths) from the central source occurs at $t=t_i$ (i.e. a Dirac delta signal in time), then the BLR emission occurs at $t=  t_i' =  t_i + \tau_{\textsc{rm}}$, where the reverberation mapping time delay is
\begin{equation}
\tau_{\textsc{rm}} (\bm{r}) = \frac{(1+z_s)}{c} \left( r - \bm{r} \cdot \hat{r}_z  \right). \label{tauBLR}
\end{equation}
Here $\hat{r}_z$ is the unit vector towards the observer and $\bm{r} =(r_x,r_y,r_z)^T = (D_s \delta \bm{\beta}, \delta \xi_z)^T$ is the spatial position of a given BLR cloud relative to the central source. The cosmological redshift factor $(1+z_s)$ corrects the time delay in the quasar rest frame for the time delay as measured in our observer frame. If we hold $\tau_{\textsc{rm}}$ fixed, then the ``iso-delay surface'' is a paraboloid of revolution around $\hat{r}_z$. Equation \eqref{tauBLR} therefore describes a family of nested paraboloids parametrised by $\tau_{\textsc{rm}}$. For the unlensed quasar, the time at which the flux from the BLR is seen is determined by the intersection of the BLR geometry with the iso-delay paraboloids.

\begin{figure}
    \centering
    \includegraphics[width=\linewidth]{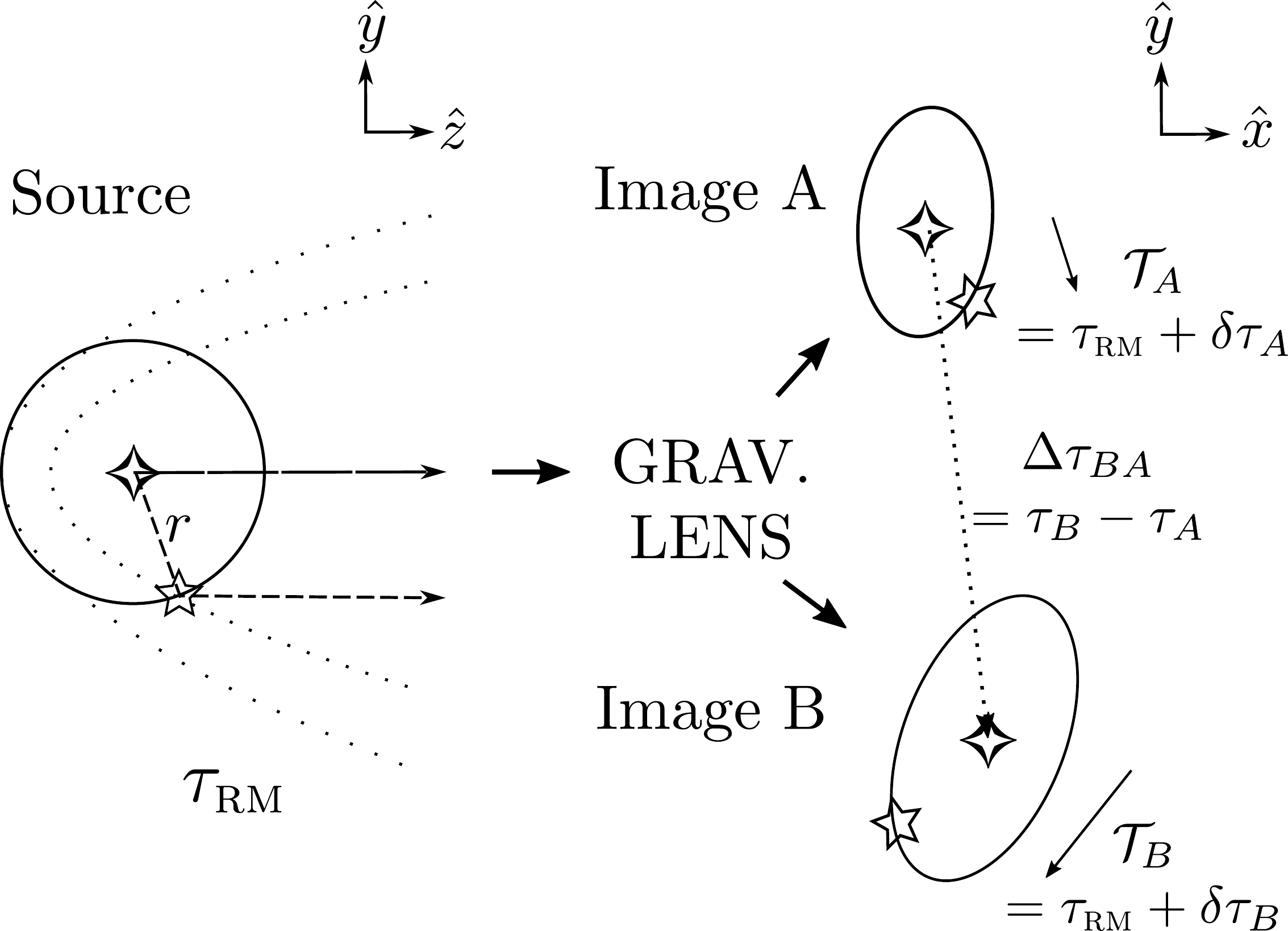}
    \caption{Schematic diagram. The source is shown projected in the y-z plane, where $z$ is in the observer direction, and the lensed images are projected in the x-y (sky) plane. The images are not spatially resolved. The central source is marked by a diamond and a chosen point on the BLR is marked by a star; their corresponding locations within the lensed images are marked similarly. There is a geometric time delay $\tau_{\textsc{rm}}$ between photons from the centre of the source and those reabsorbed and re-emitted by the BLR (paths marked by the dashed arrows; representative slices of iso-delay paraboloids marked by dotted lines); and a further differential time delay arises across each image from gravitational lensing ($\delta \tau_A$ and $\delta \tau_B$).}
    \label{fig:schematicdiagram}
\end{figure}

Considering the effect of the lensing time delay, the time that we see the continuum signal, originating from the centre of the source at $\bm{\beta}$, in image X is given by
\begin{equation}
    t_X = t_i + \tau_X. \label{timeXundashed}
\end{equation}
The time that we see a signal in image X originating from a BLR cloud at position $\bm{r}$, using the shorthand notation ${\tau}'_X \equiv \,\tau(\bm{\theta}_X + \delta \bm{\theta}_X , \bm{\beta} + \delta \bm{\beta}, \xi_z + \delta \xi_z) = \tau_X + \delta \tau_X$, is similarly given by
\begin{equation}
    t_X' = t_i' + \tau_X' = t_i + \tau_{\textsc{rm}} ( \bm{r})+ \tau_X + \delta \tau_X( \bm{r}). \label{timeXdashed}
\end{equation}
The differential lensing time delay within image X, may be found via a Taylor expansion to first order in $\delta \bm{\theta}_X$ and $\delta \bm{\beta}$ (the term corresponding to a line of sight displacement $\delta \xi_z$ may be omitted as it is many orders of magnitude less than that corresponding to a shift in the source plane; see \citet{Ng2020, Tie2018}), using the lens equation $\bm{\nabla}_{\bm{\theta}} \psi = \bm{\alpha}_X  = \bm{\theta}_X - \bm{\beta}$  where $\bm{\alpha}_X \equiv \bm{\alpha} (\bm{\theta}_X)$ is the scaled deflection angle, as
\begin{equation}
    \delta \tau_X (\bm{r}) = \frac{D_d}{D_{ds}c}(1+z_d)(\bm{\beta} - \bm{\theta}_X)\cdot (r_x, r_y)^T. \label{DeltaTX}
\end{equation}

The interval $\mathcal{T}_X(\bm{r})$ between the signal from a BLR cloud at $\bm{r}$ and the continuum from within a given image X is given by
\begin{equation}
\begin{split}
\mathcal{T}_X (\bm{r}) &\equiv t_X' - t_X = \tau_{\textsc{rm}} (\bm{r}) + \delta \tau_X (\bm{r})\\ 
&= \frac{(1+z_s)}{c} \left( r - \bm{r} \cdot \bm{e}_X \right) \label{quadricsurface}
\end{split}
\end{equation}
where $\bm{e}_X \equiv \left(\frac{(1+z_d)}{(1+z_s)} \hat{\alpha}_{X,x}, \frac{(1+z_d)}{(1+z_s)} \hat{\alpha}_{X,y}, 1 \right)^T$ and the physical deflection angle is $\hat{\bm{\alpha}}_X \equiv \hat{\bm{\alpha}} ( D_d \bm{\theta}_X ) = \tfrac{D_d}{D_{ds}} \bm{\alpha}(\bm{\theta}_X )$. The magnitude of $\bm{e}_X$ is the eccentricity $e_X$ of the conic section which generates the quadric surface of revolution described by Equation \eqref{quadricsurface}. We regain the unlensed case, i.e. $\tau_{\textsc{rm}}$, simply by setting the deflection angle to $0$. The effect of differential lensing is therefore to deform each iso-delay paraboloid into a one of the sheets of a two-sheeted hyperboloid of revolution about the $\hat{e}_X$ axis, since $e_X >1$ when the quasar is lensed and $e_X=1$ when the quasar is unlensed. For the lensed quasar, the time at which the flux for the BLR is seen is determined by the intersection of the BLR geometry with a family of iso-delay hyperboloids.

The difference in the differential lensing time delays of an image pair is independent of $\bm{\beta}$ and $\psi(\bm{\theta}_X)$ to first order in $\delta \bm{\beta}$ and $\delta \bm{\theta}_X$:
\begin{equation}
\begin{split}
\mathcal{T}_{BA} (\bm{r}) &= \delta \tau_{BA} (\bm{r}) \\
 &=  \frac{D_d}{D_{ds}c}(1+z_d) \left( \bm{\theta}_A - \bm{\theta}_B \right) \cdot (r_x, r_y)^T. \label{timedelaydifferenceequation}
\end{split}
\end{equation}
In general, we will write $\mathcal{T}_{BA} (\bm{r}) \equiv \mathcal{T}_B (\bm{r})  - \mathcal{T}_A (\bm{r})$ or $ \delta \tau_{BA} (\bm{r}) \equiv \delta \tau_B (\bm{r}) - \delta \tau_A (\bm{r})$ explicitly as a difference to emphasise that the time delay difference involves two separate measurements which are both functions of position. \citet{Yonehara1999, Yonehara2003} proposed and \citet{Goicoechea2002} investigated constraining the locations of spatially separated flares within gravitationally lensed quasars via measurements of this time delay difference.

\section{Proposed Method and Difficulties} \label{sec:methodanddifficulties}

The time delay difference between corresponding photometric signals in images A and B is solely determined by cosmology through the lensing distance, the geometry of the lensing configuration and the spatial separation within the source in the plane of the sky. This unusual property inspired the key proposal of paper I: to extract cosmological information via the ratio of angular diameter distances
\begin{equation}
    \frac{D_d}{D_{ds}} = \frac{c\left( \mathcal{T}_B (\bm{r})  - \mathcal{T}_A (\bm{r}) \right)}{ (1+z_d) \left( \bm{\theta}_A - \bm{\theta}_B \right) \cdot (r_x, r_y)^T} \label{cosmologyeqn}
\end{equation}
whilst alleviating the systematic uncertainties associated with lens modelling in using time delay measurements. The lens redshift and image positions can be measured directly. The time delay within an image $\mathcal{T}_X$ is found by subtracting the time $t_X$ we see a particular signal in the continuum flux, from the time $t'_X$ at which we see the corresponding signal in the flux of a broad emission line originating from a responding BLR cloud at $\bm{r}$.

 The technique as presented, centred on the usage of reverberation mapping to determine the position of the BLR cloud $\bm{r}$ responding at a given measured time delay in each image $A$ and $B$, is ill-motivated rather than one with limitations which could be directly mitigated by future improvements in observations. We summarise the interlinked difficulties as follows
\begin{enumerate}
    \item Reverberation mapping only gives general constraints on the geometric structure of the BLR. It does not uniquely map the location of individual points in the BLR, meaning that Equation \eqref{cosmologyeqn} is underdetermined.
    
    \item The projection of the BLR cloud position on the image-image axis can not be determined from reverberation mapping, but requires e.g. spatially resolving each image.
    
    \item In the general case $\tau_{\textsc{rm}} \gg | \delta \tau_{AB} |$; the flux from each image is largely insensitive to differential lensing.

    \item The determination of the geometric parameters is not a priori independent of the source position.
\end{enumerate}

A main challenge of exploiting Equation \eqref{cosmologyeqn} lies in the task of pinpointing the position of the BLR cloud responding at a given measured time delay in each image. This position also enters separately in the dot product in the denominator of the same equation, i.e. requiring the projection of the BLR cloud position on the image $A$ - image $B$ axis on the sky, which assuming a spatially unresolved BLR is not directly measurable. For example, \citet{Goicoechea2002} determined the position of a flare only along the image-image direction for a given image pair.

A key assumption of \citet{Yonehara2003} was that these flares are discrete, spatially separated and not physically correlated. If there is no physical correlation between discrete flares, then the shape of the flux arising from each of the flares is unique. Together with either lack of or minimal superposition of the flux, this allows each flare to be identified between images.

This assumption is violated when considering reverberation mapping of the BLR, which does not pinpoint the location from discrete flares. Rather, reverberation mapping utilises the response of an \textit{extended region} to the \textit{same} continuum signal to give general constraints on the BLR structure and kinematics in terms of geometric parameters. Determining the time at which we see the response from a particular position in a given image, as required by Equation \eqref{cosmologyeqn}, is in general an underdetermined problem as the response is a superposition of spectral flux from over the entire BLR.

The second consequence of the assumption that flares are not physically correlated is that the time between the discrete flares, denoted by $\Delta t_{\mathrm{burst}}$ in \citet{Yonehara1999}, in general can be arbitrarily small whilst their spatial separation is arbitrarily large.  This allows the fulfilment of the condition $\Delta t_{\mathrm{burst}} \lesssim |\delta \tau_{AB}|$ required for the time delay difference to have an appreciable effect. Choosing to identify $\Delta t_{\mathrm{burst}}$ as a characteristic difference in $\tau_{\textsc{rm}}$ between points on the BLR, this condition was fulfilled in Paper I in the special case of a face-on thin ring BLR geometry which responds simultaneously, but no longer holds for a general configuration in which the difference in $\tau_{\textsc{rm}}$ is on the scale of the light crossing time over the BLR structure. In the general case $\tau_{\textsc{rm}} \gg | \delta \tau_{AB} |$ and so the dependence of the flux from each image on differential lensing is very small.

To see this clearly, in the time delay within the image $\mathcal{T}_X$ given by Equation \eqref{quadricsurface} the terms $\frac{D_d}{D_{ds}}$, $(1+z_s)$ and $(1+z_d)$ are all on the order of unity, whereas $\alpha_{X}$ is on the order of at a minimum $10^{-6}$ to a maximum of $10^{-3}$ radians considering galaxy scale to cluster scale lensing. That is, in general the time scale for the differential lensing (minutes to days, depending on the lens mass) is very small compared to the time scale for reverberation mapping (tens to hundreds of days): $\delta \tau_X \ll \tau_{\textsc{rm}}$ implying $\mathcal{T}_X \sim \tau_{\textsc{rm}}$.

Finally, the determination of the geometric parameters using reverberation mapping is also not a priori independent of the distance ratio $\frac{D_d}{D_{ds}}$ nor the source position $\bm{\beta}$. We show in Section \ref{sec:parameterestimation} of the present paper however that geometric parameters may be inferred accurately without either further data (e.g. image of the Einstein ring) or lens modelling.

As an additional note, \citet{Tie2018} showed in the context of delayed emission across an accretion disk,  whilst the differential lensing time delay given by \eqref{DeltaTX} is negligible, the terms equivalent to the reverberation mapping terms in this work no longer cancel between images when considering the effect of microlensing. Rather, differential magnification over the source is included as an extra weighting factor to these terms which is image and time dependent (as the source moves relative to the stars causing the microlensing). Microlensing thereby adds time delays on the scale of the light crossing time of the relevant quasar structure to the standard time delay \eqref{tauBA}, potentially causing systematic problems for time delay cosmography; although no evidence for this effect has been observed from current data \citep[e.g.][]{Wong2020}. We do not explore this microlensing effect as it does not impact the above conclusions centring on the potential-free expression for the differential lensing time delay.

\subsection{The Degeneracy Problem Exemplified}

We now illustrate the problem of underdetermination with an example. The spectral flux from axisymmetric BLR geometries including the thick ring or disk, the spherically symmetric and the biconical geometry may be considered as the superposition of the contribution of the spectral flux from (possibly inclined) thin rings. We therefore write the Cartesian BLR body coordinates $\bm{p} = r (\sin \vartheta \cos \varphi, \sin \vartheta \sin \varphi, \cos \vartheta)^T$ in terms of spherical coordinates where $\vartheta$ is the zenith angle and $\varphi$ is the azimuthal angle, which is related to the coordinates $\bm{r}$ (which we defined such that $\hat{r}_z$ is in the observer, or line-of-sight, direction) by a rotation through an inclination angle $i \in [ 0 , 2 \pi)$ about the $\hat{r}_x$ axis such that $\bm{r} = \text{Rot}_x(i) \bm{p}$. For example, a ring or disk co-planar with the central ionising source has fixed $\vartheta = \frac{\pi}{2}$, and when its inclination angle is $i=0$ it is face-on and when $i = \frac{\pi}{2}$ it is edge-on with respect to the observer. A spherical geometry may have $i$ set to be $0$.

To demonstrate the degeneracy issue, we restrict Equation \eqref{cosmologyeqn} to the least-degenerate case of an infinitesimally thin inclined ring such that the position within the BLR geometry is parametrised by $\varphi$ alone. 
\begin{equation}
    \frac{D_d}{D_{ds}} = \frac{c \left(\mathcal{T}_B (R, \tfrac{\pi}{2}, \varphi)  - \mathcal{T}_A (R, \tfrac{\pi}{2}, \varphi) \right)}{ R (1+z_d) \left(  \theta_{AB,x} \cos \varphi + \theta_{AB, y} \cos i \sin \varphi \right)} \label{cosmologyeqn_thinring}
\end{equation}
which assuming circular Keplerian orbital motion can be related to the line-of-sight velocity variable $v_z$ (and thereby the observed wavelength) via a line-of-sight velocity model $u_z(\bm{r})$:
\begin{equation}
    u_z (r, \varphi)= u(r) \sin i \cos \varphi
\end{equation}
where $u(r) = \sqrt{\tfrac{GM}{r}}$ and $M$ is the central black hole mass. So for each value of the line of sight velocity, there are two corresponding positions $\varphi^+$ and $\varphi^-$ which are defined by
\begin{equation}
    \sin \varphi^\pm \equiv \pm \sqrt{1 - \left( \tfrac{v_z}{u(R) \sin i} \right)^2}
\end{equation}
such that $0 \leq \varphi^+ \leq \pi$ and $\pi \leq \varphi^- \leq 2 \pi$. One can immediately see that the value of the denominator of Equation \eqref{cosmologyeqn_thinring} is dependent on the choice of $\varphi^{\pm}$ and $\mathcal{T}_{BA} ( \varphi^+ ) \neq \mathcal{T}_{BA} ( \varphi^-)$: even in the simplest case of an infinitesimally thin ring, the positions are degenerate. In Paper I we disregarded this position-wavelength degeneracy (all positions correspond to a single wavelength for a face-on ring) and further exploited the symmetry of the projection of a thin face-on ring in the $\textit{x-y}$ plane: the dependence on position $\varphi$ within the BLR geometry disappears in the denominator when taking the maximum of the time delay for the special case of a face-on ring.

In more realistic models, such as a radially-thick inclined ring, we no longer have a fixed $R$ but rather a range of values of $r$ (and also of $\cos \varphi^\pm$). Each $\mathcal{T}_X(r, \tfrac{\pi}{2}, \varphi^\pm)$ term as well as $\sin \varphi^\pm$ now corresponds to \textit{two} ranges of values; the system is underdetermined and we can not find constraints on $\mathcal{T}_B (\bm{r}) - \mathcal{T}_A (\bm{r})$ for a given inferred value of $v_z$.

\section{Combining Differential Lensing and Reverberation Mapping} \label{sec:velocitydelaymaps}

Reverberation mapping provides constraints on the overall geometry of the BLR since measurements of $T_X$ provides information on $\bm{r}$; and in addition, the observed wavelength within the broad emission line is proportional (assuming relativistic effects are negligible\footnote{A full analysis would consider the impact of relativistic effects which are significant for small BLR radii, such as gravitational redshifting, the relativistic Doppler effect and relativistic beaming, e.g. \citet{Corbin1997, Goad2012}. A consequence of the relativistic Doppler effect is that in general the observed wavelength is not simply proportional to line-of-sight velocity.})
to the line-of-sight velocity of the BLR cloud. If the velocity of a BLR cloud may be related to its position  via a model for the velocity field, then the line-of-sight velocity gives additional information on the position of the BLR cloud as well as the kinematics. A  ``cloud'' refers to small individual entities in the emission line region in reverberation mapping literature; we interchangeably refer to either a cloud or a point particle of the BLR distribution.

We first review the unlensed case \citep{Blandford1982, Peterson1993, Netzer1990, Netzer2013}. Let $\bm{w}$ be the velocity coordinates of a BLR cloud, whereas $v_z$ denotes the line-of-sight velocity variable. The broad emission line spectral flux responds to the continuum flux $C(t - \tau_{\textsc{rm}})$, as given in generality by
\begin{equation}
    L( v_z , t) = \iint j(\bm{r}) C ( t' - \tfrac{r}{c} ) g (\bm{r}, v_z) \delta ( t - (t' - \tfrac{r}{c} + \tau_{\textsc{rm}}(\bm{r}))) d \bm{r} d t'
\end{equation}
where the line-of-sight projected 1D velocity distribution $g(\bm{r}, v_z)$ is defined in terms of the full velocity distribution $f( \bm{r}, \bm{w})$ as
\begin{equation}
    g(\bm{r}, v_z) \equiv \int f( \bm{r}, \bm{w}) \delta ( v_z - \bm{w} \cdot \hat{r}_z ) d \bm{w}.
\end{equation}
The responsivity term $j(\bm{r})$ is in general dependent on the position (or if isotropic, simply the radius) of the cloud and contains the physics of the photoionisation of the BLR. The continuum flux received by a gas cloud at time $t'$ and position $\bm{r}$, emitted by the central source at an earlier time $t' - \tfrac{r}{c}$, is $j(\bm{r}) C ( t' - \tfrac{r}{c} ) = \frac{\varepsilon(\bm{r})}{4 \pi r^2} C ( t' - \tfrac{r}{c} )$ and plays the role of the emissivity of the cloud where $\varepsilon(\bm{r})$ is the reprocessing coefficient of the cloud.

The general reverberation mapping problem involves finding the deconvolution of
\begin{equation}
    L (v_z, t) = \int \Psi(v_z, s) C (t-s) ds
\end{equation}
since all of the geometrical information is contained within the transfer function $\Psi (v_z, t)$. When the transfer function is visualised using a heat map in $(v_z, t)$-space, it is commonly referred to as a ``velocity-delay map''. This deconvolution problem may be approached using maximum entropy fitting techniques, or regularised linear inversion; and then the outputted velocity-delay map may be compared qualitatively to the theoretical velocity-delay maps produced by simple models. Recovering the transfer function and thereby inferring the physical and kinematic distribution of the BLR using reverberation mapping has been, up until recent improvements of the quality of reverberation data sets, regarded as a technique in a developmental state \citep{Shen2015, Mangham2019, Cackett2021}. Rather than solving this ill-posed inverse problem, reverberation mapping has been often limited to finding only the mean radius for the BLR through measuring the mean time delay for emission lines using cross-correlation analyses, from which it is possible to constrain the mass of the central black hole. Forward modelling of the BLR also offers an alternative Bayesian approach to directly (without explicitly finding the transfer function) providing best-fit parameters of a flexible BLR geometry \citep{Pancoast2011}. 

The transfer function, or the Green's function for the system, is the line response to a Dirac-delta continuum pulse; replacing $C (t' - \tfrac{r}{c})$ with $\delta (t' - \tfrac{r}{c})$ gives
\begin{align}
    \Psi (v_z, t) &= \int j(\bm{r}) g (\bm{r}, v_z) \delta ( t - \tau_{\textsc{rm}}(\bm{r})) d \bm{r}\\
    &= \int j(\bm{r}) n(\bm{r}) \delta (v_z - u_z(\bm{r})) \delta ( t - \tau_{\textsc{rm}}(\bm{r})) d \bm{r}.
\end{align}
The second equality holds when there exists a velocity model $\bm{u}(\bm{r})$ whose corresponding line-of-sight velocity model is $u_z(\bm{r})$, such that we may write the velocity distribution as
$f(\bm{r}, \bm{w}) = n(\bm{r}) f(\bm{w}| \bm{r}) = n(\bm{r}) \delta (\bm{w} - \bm{u}(\bm{r}))$
where $n(\bm{r})$ is the number density of responding clouds. This gives the line-of-sight projected velocity distribution as $g(\bm{r}, v_z) = n(\bm{r}) \delta (v_z - u_z(\bm{r}))$. We can interpret the general problem as one of finding a probability density function under a change of variables from positions $\bm{r}$ to $(v_z , T_X)$. As the dimensionality of the original set of random variables is not the same as the new set of random variables, the probability density functions are written using Dirac delta generalised functions.

Including the effect of differential lensing on the transfer function simply involves substituting the reverberation mapping time delay function $\tau_{\textsc{rm}}(\bm{r})$ with the time delay function $\mathcal{T}_X(\bm{r})$ and denoting the time variable with the subscripted $T_X$ to emphasise that it is measured with respect to the lensed continuum signal in each image $X$. It follows that the transfer function and the observed broad emission line spectral flux of each image are respectively given by
\begin{equation}
    \Psi_X (v_z, T_X) = \int j(\bm{r}) n(\bm{r}) \delta (v_z - u_z(\bm{r})) \delta ( T_X - \mathcal{T}_X(\bm{r})) d \bm{r} \label{eq:spectralflux}
\end{equation}
and
\begin{equation}
    L_X (v_z, T_X) = \int \Psi_X(v_z, s) C (T_X -s) ds. \label{lensedlineresponse}
\end{equation}

Although overall deductions about the BLR geometry and kinematics made be made from finding the transfer function or else by direct forward modelling, determining the time at which we see the response from a particular position in a given image, as is required by Equation \eqref{cosmologyeqn}, is in general an underdetermined problem. For the simple BLR models usually considered, e.g. where the (emissivity-weighted) number density distribution is modelled as uniform over the line, surface or volume to which the particles are confined and does not add an extra parameter, the full transfer function does not give information further than its support. For inhomogeneous BLR models which feature over- or under-densities of clouds in particular spatial regions however, the form of the transfer function may further constrain the number density distribution over that region. It can still be somewhat useful to find the theoretical transfer functions for idealised models for the visualisation of the method. 

We therefore present the form of the transfer function in the case of a uniform thin inclined co-planar ring in Section \ref{subsec:inclinedringdisk} below, and leave the details as well as calculations for a disk and spherical shell to Appendix \ref{appendixa}. Throughout these calculations, we assume $j(\bm{r})$ to be a constant \citep[e.g.][]{Pancoast2011}, i.e. that the radiation received and re-emitted by all BLR clouds is constant. A more physically accurate assumption of the dependence of $j(\bm{r})$ on, for example, radius does not impact the overall findings discussed in Section \ref{sec:methodanddifficulties} of this paper. As light is simply redistributed in time due to the differential lensing, the overall effect on the spectral flux tends to be a small change in magnitude of the transfer function whilst spanning a correspondingly larger or smaller domain in time compared to reverberation mapping without lensing.

\subsection{General Effects of Differential Lensing on The Shape of The Velocity Delay Map} 

We can distinguish the time at which the signal from some subset of the clouds, in image A from the time which it is seen in image B, by comparing the support of the transfer function in $(v_z,T_X)$-space from different images; that is, analysing the effect of differential lensing on the shape of the velocity-delay map. This is dictated purely by the first two of the available constraints, the time delay $\mathcal{T}_X(\bm{r})$ and the velocity model $u_z (\bm{r})$.

In this work, we consider axisymmetric broad line region geometries, with three classes as explicit examples: the ring or disk, the spherically symmetric, and the biconical geometry as examples. For non-axisymmetric geometries, the overall principles should still apply, but the analysis may not be as straightforward. It is useful to variously express the time delay within an image $\mathcal{T}_X$, Equation \eqref{quadricsurface}, in spherical coordinates as
\begin{subequations}
\begin{align}
\begin{split}
    &\mathcal{T}_X (\bm{r}) \\
    &= \tfrac{r}{c} \left(1+z_s + J_X \cos \vartheta +  A_X \sin \vartheta  \sin \varphi + B_X \sin \vartheta\cos \varphi\right) \label{timedelayequation_zero}
\end{split}\\
    &= \tfrac{r}{c} \left( 1+z_s + J_X \cos \vartheta + F_X \sin \vartheta \sin ( \varphi + \phi_X) \right) \label{timedelayequation_one} \\
    &= \tfrac{r}{c} \left( 1+z_s + J_X \cos \vartheta + F_X \sin \vartheta \sin ( \mathrm{sgn}(A_X) \varphi + \phi_X^+) \right) \label{timedelayequation_two}
\end{align}
\end{subequations}
by defining the parameters:
\begin{subequations}
\begin{align}
   J_X &\equiv - (1+z_d) \hat{\alpha}_{X,y} \sin i - (1+z_s) \cos i \\
   A_X &\equiv (1+z_s) \sin i - (1+z_d) \hat{\alpha}_{X, y} \cos i\\
   B_X &\equiv  -(1+z_d) \hat{\alpha}_{X,x}\\
   F_X &\equiv \sqrt{A_X^2 + B_X^2}\\
   \phi_X &\equiv \begin{cases}
    \phi_X^+ \equiv  \arcsin{\frac{B_X}{F_X}} &  A_X \geq 0\\
   \pi - \phi_X^+ &  A_X < 0,\; B_X \geq 0\\
   -\pi - \phi_X^+ &  A_X < 0,\; B_X <0\\ 
   \end{cases}
\end{align}
\end{subequations}
where $\phi_X \in [-\pi, \pi)$ solves $\cos \phi_X = \frac{A_X}{F_X}$ and $\sin \phi_X = \frac{B_X}{F_X}$ simultaneously, and we set $\mathrm{sgn}(0) =1$. The functional forms of $\tau_{\textsc{rm}} (\bm{r})$ and $T_X (\bm{r})$ are identical; we can regain $\tau_{\textsc{rm}}$ by setting the deflection angle $\hat{\alpha}_{X}$ to $0$ such that $J_X = -(1+z_s) \cos i$, $F_X = (1+z_s) \sin i$ and $\phi_X^+ = 0$.

We consider rotational and radial velocity fields whose line-of-sight components are given respectively by
\begin{align}
    u_{\mathrm{Rot}, z } (\bm{r}) &= u(r) \sin i_v \sin \vartheta \cos \varphi \label{rotationalmotioneqn}\\
    u_{\mathrm{Rad}, z} (\bm{r}) &= u(r) (\cos i \cos \vartheta - \sin i \sin \vartheta \sin \varphi ) \label{radialmotioneqn}
    \end{align}
where the rotation axis is in the direction given by $\text{Rot}_x(-i_v) \hat{r}_z$, and $i_v = i$ for disk and biconical geometries and $i_v \in [0, 2 \pi)$ is chosen independent of fixing $i = 0$ for spherical geometries. In the case of ring or disk geometries, we consider circular Keplerian orbits (Equation \eqref{rotationalmotioneqn} with $u(r) = \sqrt{\tfrac{GM}{r}}$ and $\vartheta = \frac{\pi}{2}$) whereas in the case of spherically symmetric geometries we consider solid body rotation (Equation \eqref{rotationalmotioneqn} with $u(r) \propto r$). Combining the time-delay constraint \eqref{timedelayequation_one} with either velocity model \eqref{rotationalmotioneqn} or \eqref{radialmotioneqn} gives a pair of parametric equations for the support of the velocity-delay map. 

\begin{figure}
    \centering
    \includegraphics[width=\linewidth]{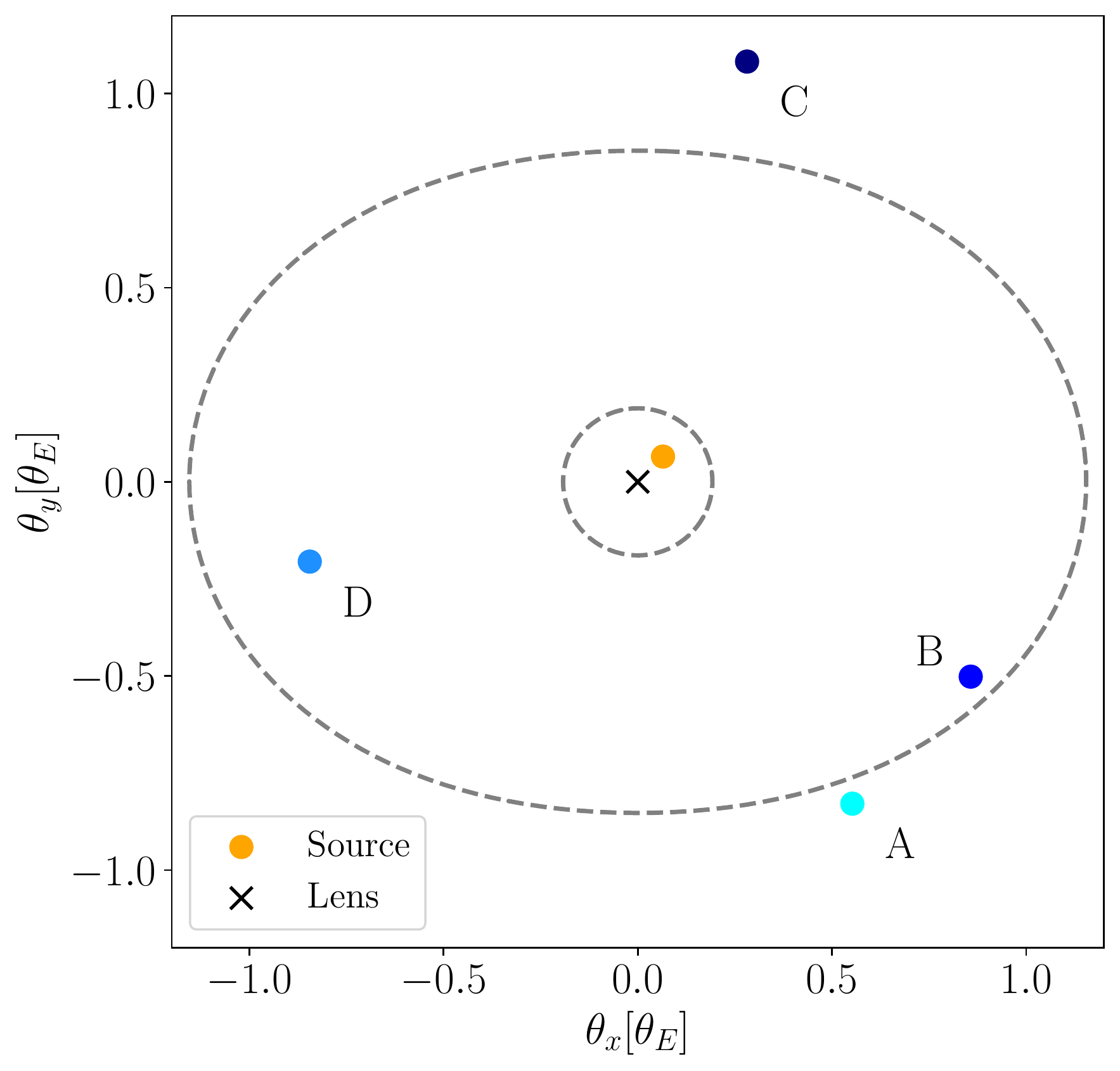}
    \caption{We reproduce the lensing configuration shown in Figure 3 of Paper I, which is based on a softened elliptical lens model with a core radius of $0.1 \theta_E$ and ellipticity of $0.1$, setting the source at $\bm{\beta} = (0.065, 0.065) \theta_E$, where $\theta_E$ denotes the deflection scale of the lens. The critical lines are marked by dashed grey lines. The calculated image positions are used for the illustrative examples throughout this work.}
    \label{fig:sieconfig}
\end{figure}

\begin{figure}
    \centering
    \includegraphics[width=\linewidth]{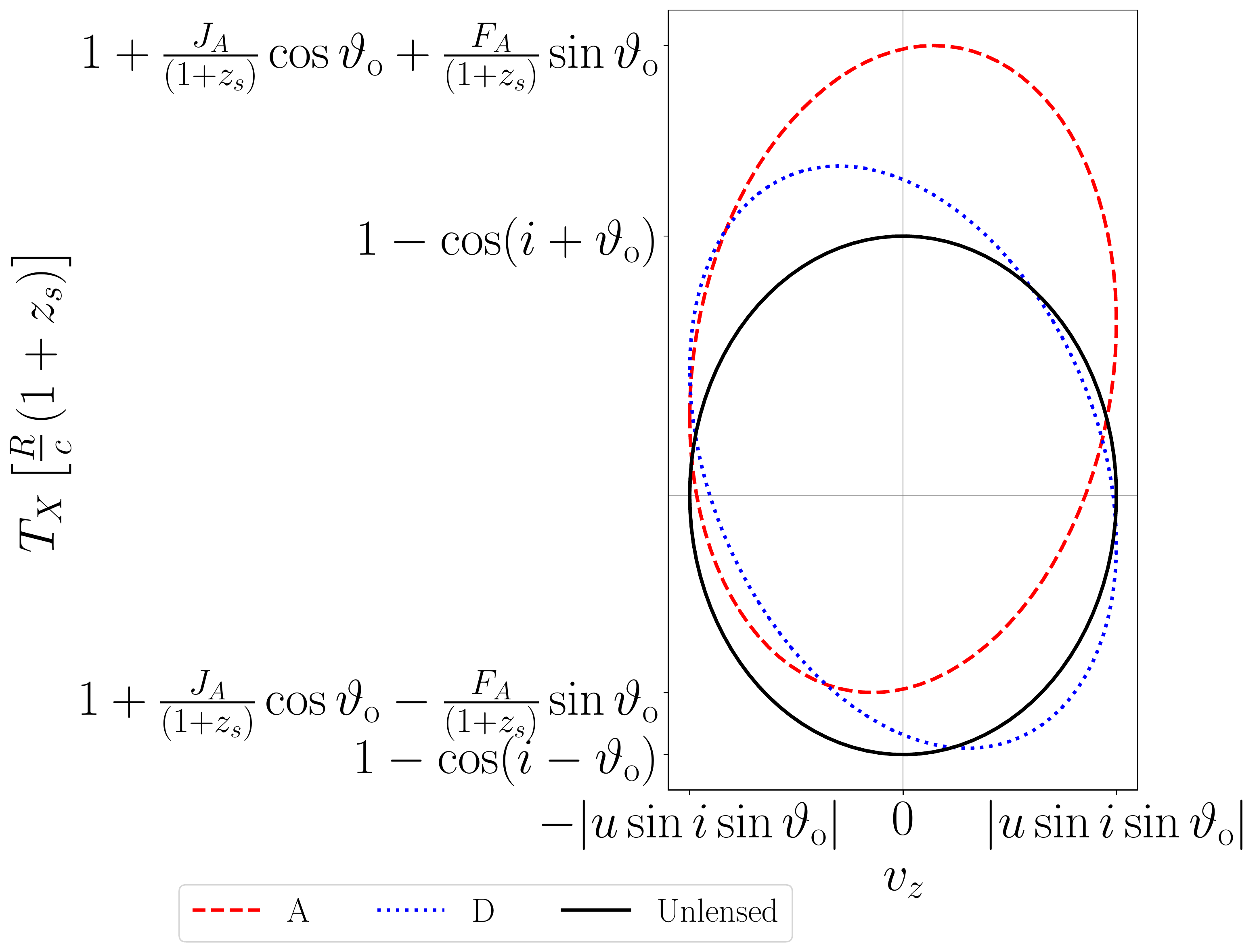}
    \caption{The general effect of differential lensing on the velocity-delay map of a single thin inclined ring component of a BLR geometry in solid body rotation, where the radius of the ring is $r = R$, at fixed polar coordinate $\vartheta_\mathrm{o}$ (i.e. not co-planar with the central source) and with inclination angle $i$. The effect of the lensing is exaggerated by choosing the scaled deflection $\alpha_X$ to be on the order of $10^{-2}$ radians, such that we may distinguish between the ellipses over the entire velocity-delay map for the purpose of this illustration. The observed ellipse is shifted and re-scaled in the time dimension, and rotated by a magnitude which differs between observed images (A and D of Figure \ref{fig:sieconfig} shown here). The axes of the unlensed ellipse are marked by solid lines.}
    \label{fig:singleringdelaymap}
\end{figure}

Recall that the spectral flux from any of these BLR geometries may be considered as the superposition of the contribution of the spectral flux from thin rings. The velocity-delay maps of these geometries may therefore be analysed by decomposing the structures into thin rings, i.e. slicing by holding $r$ and $\vartheta$ constant or alternatively (corresponding to halves of thin rings) $r$ and $\varphi$ constant. In either case, we see that a single thin ring component of a general geometry sketches out an ellipse in $(v_z, T_X)$-space. In either the lensed or unlensed case, the ellipse may be degenerate (i.e. forming a point or an interval, dependent on the parameters).

The general effect of differential lensing is to alter the eccentricity, rotate and scale the axes of, and shift in the time-delay dimension this $(v_z,T_X)$-space ellipse as illustrated in Figure \ref{fig:singleringdelaymap}. The parameter $J_X$ controls the shift in time and $F_X$ the scaling in time or the half-height of the ellipse, whereas the phase-shift $\phi^{+}_X$ controls the rotation and eccentricity. For typical values of the differential lensing time delay, there is negligible rotation of the velocity-delay ellipse for all values of the inclination angle away from the face-on case where $i=0$. For all values of $i$ away from $i=0$, the scale of the velocity-delay ellipse on the time axis is determined by the radius of the thin ring and on the scale of tens to hundreds of days whereas the effect of the lensing is on the scale of minutes. We discuss the consequences in detail for each class of BLR geometry.

\subsection{Inclined Ring or Disk} \label{subsec:inclinedringdisk}

\begin{figure}
    \centering
    \includegraphics[width=\linewidth]{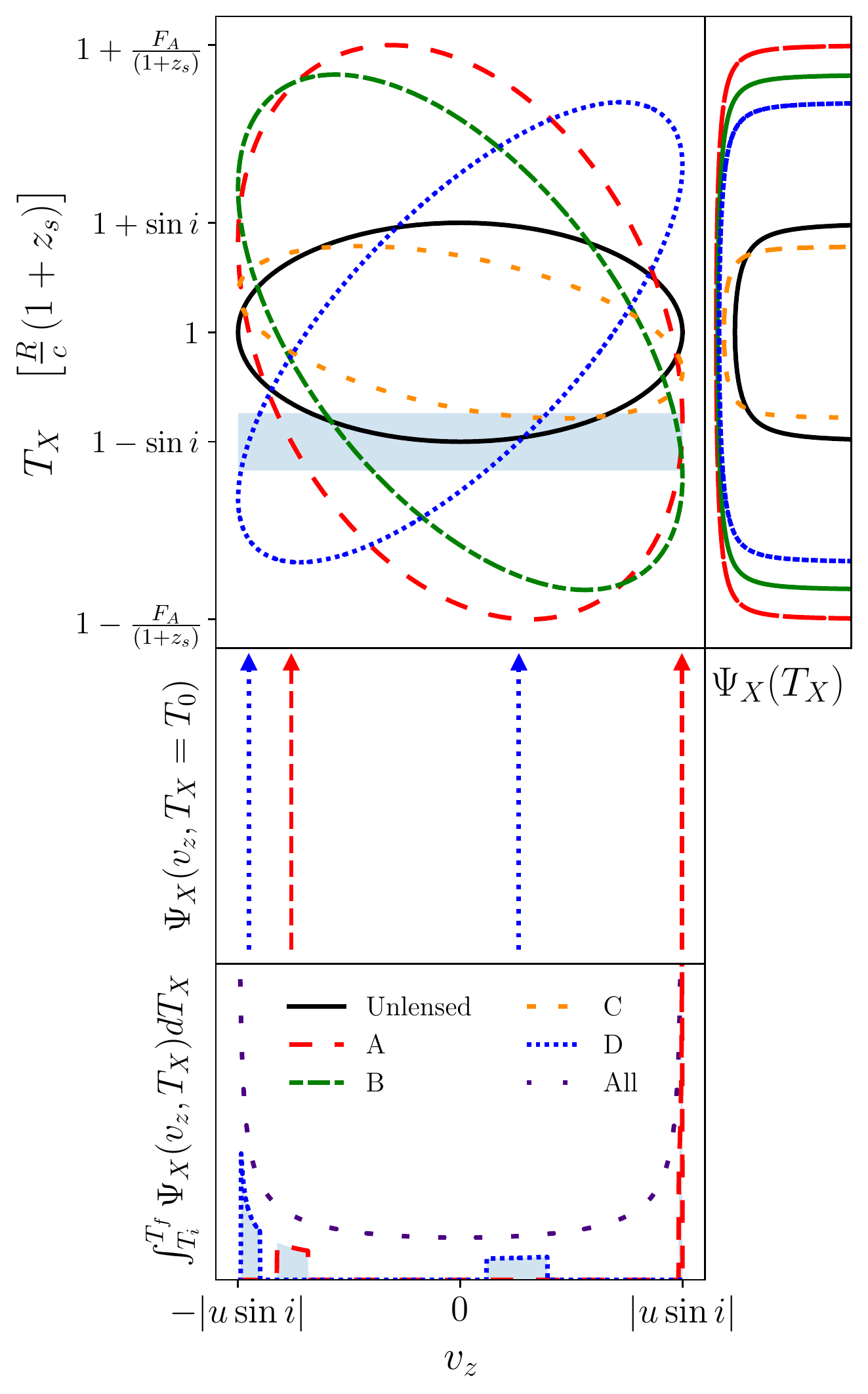}
    \caption{Velocity-delay map given by Equations \eqref{Txthinring} and \eqref{vzthinring}, transfer function integrated over all wavelengths $\Psi_X(T_X)$ given by Equation \eqref{fluxthinring} and transfer function at and over a given time $\big( \Psi_X(v_z, T_X= T_0)$ and $\int_{T_i}^{T_f} \Psi_X (v_z, T_X) d T_X \big)$ from image X $= \{$A, B, C, D$\}$ (see Figure \ref{fig:sieconfig}) of a lensed inclined thin ring BLR in Keplerian orbit. We here have chosen galaxy-scale lensing with a very small inclination angle $i= 10^{-6} \pi \, \mathrm{rad}$ for the purpose of illustration, which leads to the difference in height between the ellipses being a relatively large fraction of their overall heights, and also the relatively dramatic rotations between them. This allows the ellipses of all images and of the unlensed case to be easily distinguished in the illustration. The difference in height between ellipses as a fraction of their overall height becomes very small very quickly away from $i=0$ or $i=\pi$, as does the magnitude of their relative rotations. We show the spectral flux from images A and D at a time delay $T_0$ within images A and D are Dirac delta distributions as given by Equation \eqref{spectralfluxthinring} which when integrated over a time from $T_i = T_0 - \Delta T$ to $T_f = T_0 + \Delta T$ (shaded blue region) form segments of an arcsine distribution, as given by Equation \eqref{spectrafunc_thinring}. %We have picked a value of $T_0 = \frac{R}{c} (1+z_s) \left(1 - \sin i \right)$ and  $\Delta T = \frac{R}{c} \left(\tfrac{F_A}{20} \right)$.
    The relatively large offsets and asymmetry in the $v_z$ locations of the spectral flux spikes at this arbitrary value of $T_0$ is due to the choice of the small inclination angle. The spectral flux integrated over all time $L(v_z)$ is a complete arcsine distribution given by the dotted purple line, independent of lensing and thus identical for all images. The relative difference in the magnitude of the spectral flux when integrated over the short time interval versus all time is due to the differing contributions from the multiple roots in Equation \eqref{spectralfluxthinring}.}
    \label{fig:thinringveldelmap}
\end{figure}

Consider as the simplest model for a BLR geometry a thin ring co-planar with the continuum source with fixed $\vartheta = \frac{\pi}{2}$ and radius $r=R$, inclined at an angle $i \in [ 0 , 2 \pi)$; in Keplerian orbit with orbital speed $u$. Equations \eqref{timedelayequation_two} and \eqref{rotationalmotioneqn} are then parametrised by the azimuthal angle $\varphi \in [0, 2\pi)$, with the direction of parametrisation determined by $\mathrm{sgn}(A_X)$:
\begin{subequations}
\begin{align}
       \mathcal{T}_X (\varphi) &= \tfrac{R}{c} \left(1 +z_s  +  F_X \sin{\left(\mathrm{sgn}(A_X) \varphi + \phi_X^+  \right)} \right) \label{Txthinring}\\
        u_{z}(\varphi) &= u \sin i \cos \varphi \label{vzthinring}
\end{align}
\end{subequations}
tracing an ellipse in ($v_z, T_X$)-space as illustrated in Figure \ref{fig:thinringveldelmap}. As in the general case, lensing alters the values of $F_X$ and $\phi_X^+$ where $ \frac{R}{c}F_X$ determines the half-height of the ellipse, and the phase shift angle $\phi_X^+$ determines the rotation of the ellipse axes from the $(v_{z}, T_X)$-axes in addition to modifying the eccentricity. However, for the co-planar thin ring, there is no shift in time. 

\begin{figure*}
    \renewcommand{\thefigure}{\arabic{figure}a}
    \centering
    \includegraphics[width=\linewidth]{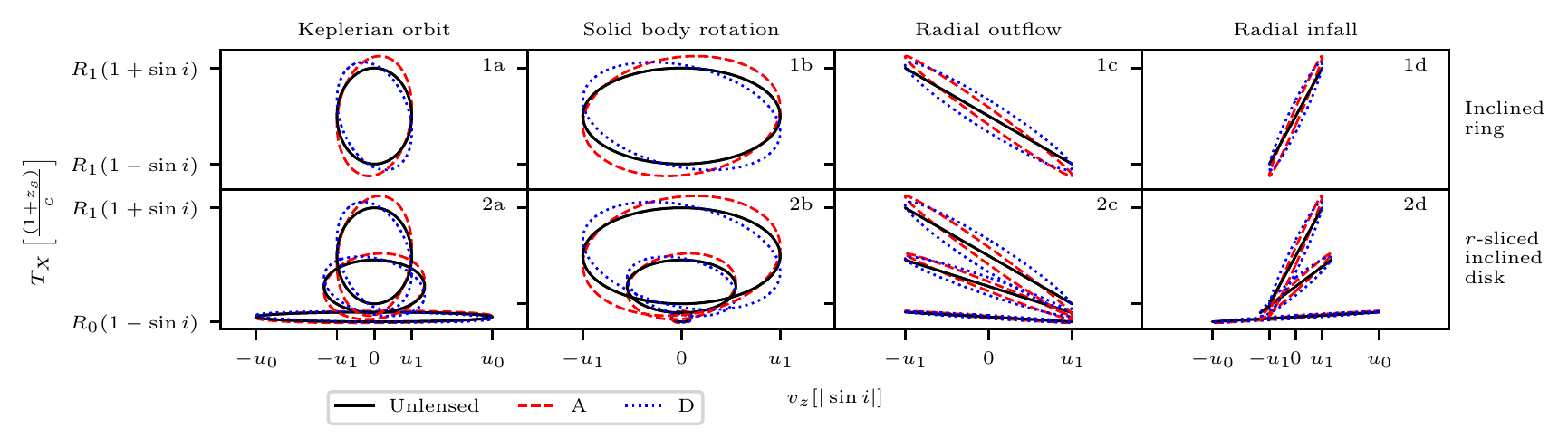}
    \caption{An illustration showing showing the effect of lensing on velocity-delay maps (the red or blue compared to the solid black lines), greatly exaggerated by choosing the scaled deflection $\alpha_X$ to be on the order of $10^{-2}$ radians. Row 1 corresponds to a thin inclined ring (co-planar with the continuum source) and row 2 to three representative radial slices at $R_0 \equiv R_{\mathrm{min}}$, $R_1 \equiv R_{\mathrm{max}}$ and $ (R_0 + R_1)/2$ of an inclined disk (also co-planar with the continuum source); both with inclination angle $i= \frac{\pi}{4}$. Column a corresponds to Keplerian orbits, column b to solid body rotation, column c to constant radial outflow and column d to radial infall, where $u_0 \equiv |u(R_\mathrm{min})|$ and $u_1 \equiv |u(R_\mathrm{max})|$.}
    \label{fig:alldelaymaps_disk}
\end{figure*}

The velocity-delay map for a radially-thick inclined ring or disk can be generated as a superposition of the velocity-delay maps for the single inclined thin ring over a range of radii: see Figure \ref{fig:alldelaymaps_disk} subplot 2a. This results in a characteristic bell shaped envelope that will be slightly deformed by the lensing. The bell shape is elongated in the time domain in accordance with the maximum and minimum radius orbits, since the top and bottom points of this bell shape corresponds to the maximum and minimum radius orbits respectively.

For the thin inclined ring BLR with uniform linear number density $\mu$, the number density distribution of the BLR model is
$n(\bm{r}) = \mu \delta(r - R) r^{-1} \delta(\vartheta - \frac{\pi}{2})$. The one-dimensional transfer function, as illustrated in Figure \ref{fig:thinringveldelmap}, is given by an arcsine distribution
\begin{equation}
    \Psi_X (T_X) \propto \frac{
    2 \mu c}{\sqrt{F_X^2 - \left(T_X c R^{-1} - (1 +z_s) \right)^2}} \label{fluxthinring}
\end{equation}
when $\left|T_X c R^{-1} - (1 +z_s) \right| \leq F_X$ and is $0$ otherwise. The two-dimensional transfer function is 
\begin{equation}
\Psi_X (v_z, T_X) \propto \mu c R \! \sum\limits_{\pm} \frac{\delta\left(T_X - W^{\pm}_X(R, v_z) \right)}{V^{+}(R, v_z)} \label{spectralfluxthinring}
\end{equation}
where 
\begin{align}
    V^{\pm}(r, v_z) &\equiv u(r) \sin i \sin \varphi^\pm = \pm \sqrt{(u(r) \sin i)^2 - v_z^2}\\
\begin{split}
    W^{\pm}_X(r, v_z) &\equiv \mathcal{T}_X (r, \tfrac{\pi}{2}, \varphi^\pm)\\
    &= \frac{r}{c} \left( 1+ z_s + A_X \frac{V^{\pm}(r, v_z)}{u(r) \sin i}  + B_X \frac{v_z}{u(r) \sin i }\right).
\end{split}
\end{align}
The summation implies that we have a Dirac delta function where $T_X = W^{+}_X(R, v_z)$ and another where $T_X = W^{-}_X(R, v_z)$, as illustrated by Figure \ref{fig:thinringveldelmap}. We note that the form of the transfer function is independent of the image position. When the two-dimensional transfer function is integrated over an arbitrary amount of time, we see that the resulting spectra is given by segments of an arcsine distribution in $v_z$, where the widths and positions of these segments are determined by the integration time,
\begin{equation}
\int\limits_{T_{X,i}}^{T_{X,f}} \Psi_X (v_z, T_X) \, dT_X \propto
 \frac{\mu c R}{V^{+}(R, v_z)} \label{spectrafunc_thinring}
\end{equation}
if $T_{X,i} \leq W^{\pm}_X(R, v_z) \leq T_{X,f}$ and is $0$ otherwise, as shown in Figure \ref{fig:thinringveldelmap}.

\subsection{Spherical Shell}
We can straightforwardly build up the velocity-delay map of a thin spherical shell geometry, which has fixed radius $r=R$, $\vartheta \in [0, \pi]$, $\varphi \in [0, 2\pi)$ and we may set the inclination angle $i=0$. If the spherical shell is in solid body rotation about a rotation axis, which is in the direction given by $\text{Rot}_x(-i_v) \hat{r}_z$ where $i_v \in [0, 2 \pi)$, the constraints become
\begin{subequations}
\begin{align}
    \mathcal{T}_X(\vartheta, \varphi) &= \tfrac{R}{c} \left( (1+z_s) (1- \cos \vartheta ) + F_X \sin \vartheta \sin (\varphi + \phi_X) \right) \label{thinsphericalshelltd}\\
    u_z(\vartheta, \varphi) &= u(R) \sin i_v \sin \vartheta \cos \varphi \label{solidbodyrotation}
\end{align}
\end{subequations}
where here $F_X$ reduces to simply $(1+z_d) \hat{\alpha}_X$.

\begin{figure*}
    \renewcommand{\thefigure}{5b}
    \centering
    \includegraphics[width=\linewidth]{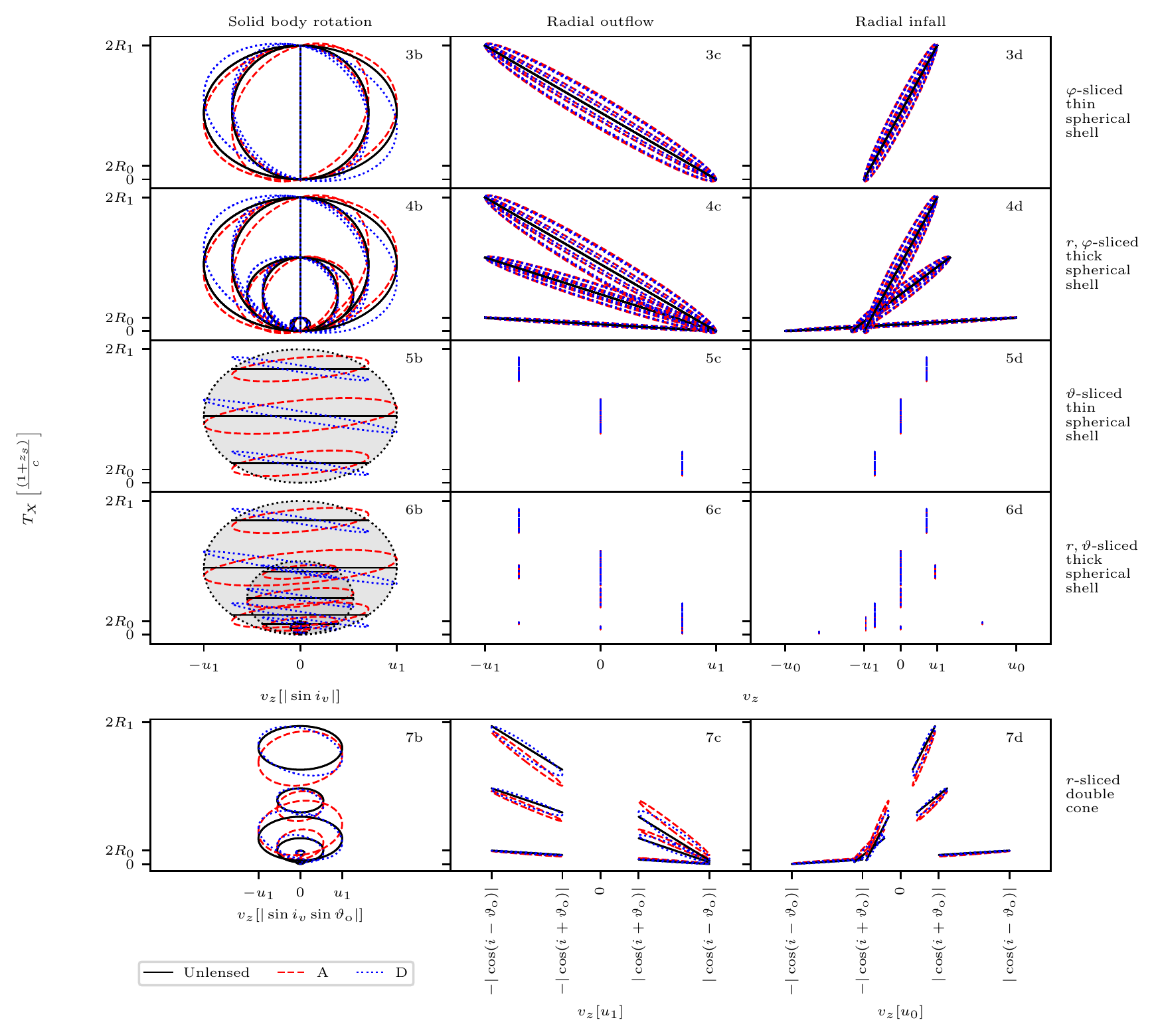}
    \caption{An illustration showing showing the effect of lensing on velocity-delay maps (the red or blue compared to the solid black lines), greatly exaggerated by choosing the scaled deflection $\alpha_X$ to be on the order of $10^{-2}$ radians. The BLR geometry is varied by row and the velocity field is varied by column as in, and labelled following on from, Figure \ref{fig:alldelaymaps_disk}. Column b corresponds to solid body rotation, column c to constant radial outflow and column d to radial infall, where $u_0 \equiv |u(R_\mathrm{min})|$ and $u_1 \equiv |u(R_\mathrm{max})|$. Rows 3 and 5 correspond respectively to representative $\varphi$-slices and $\vartheta$-slices of a thin spherical shell; and rows 4 and 6 respectively correspond to representative radial and $\varphi$-slices, and radial and $\vartheta$-slices of a thick spherical shell. Row 7 corresponds to representative radial slices of a hollow bicone with half-opening-angle $\vartheta_\mathrm{o} = \frac{\pi}{7}$. All radial slices are taken at $R_0 \equiv R_{\mathrm{min}}$, $R_1 \equiv R_{\mathrm{max}}$ and $ (R_0 + R_1)/2$; five $\vartheta$-slices are taken at equal intervals between $0$ and $\pi$ and nine $\varphi$-slices taken at equal intervals between $0$ and $2\pi$. In 3b and 4b, whilst the ellipses corresponding to $\varphi = (\pi +) \frac{\pi}{4}$ and $(\pi +) \frac{3\pi}{4}$ are degenerate in the unlensed case, they are not when lensed. Considering instead $\vartheta$-slices, 5b and 6b illustrate that when unlensed, each slice draws out an interval in ($v_z, T_X$)-space (black solid lines), such that the complete unlensed velocity-delay map is a filled ellipse, the envelope of which is shown by the black dotted line. Each of these intervals are distorted into ellipses (red and blue dashed lines) as shown, with the largest distortion in time at $\vartheta = \frac{\pi}{2}$. When the overall structure is considered, this lensing effect is obscured by the contributions from each thin slice. For radial motion, slicing by $\vartheta$ gives vertical intervals at each $\vartheta$ value: showing more $\vartheta$ values would fill out into the ellipses apparent in rows 3 and 4. For the biconical geometry in solid body rotation in 7b, each radial slice of the biconical structure gives a thin ring at $\vartheta = \vartheta_\mathrm{o}$ and another at $\pi - \vartheta_\mathrm{o}$; hence we see three pairs of ellipses. Considering instead radial motion and slicing by $\varphi$ (3c, 3d, 4c and 4d), each thin spherical shell when unlensed maps to diagonal intervals (black) where spherical outflow corresponds to a negative slope and positive for infall (as the observer is in the $+ \hat{z}$ direction). When lensed, these intervals distort into the blue and red filled ellipses. The bicone, as a subset of a thick spherical shell, has its velocity-delay map as a subset of that of the thick spherical shell: i.e 7c is a subset of 4c and 7d of 4d, dependent on the inclination angle $i$ and half-opening angle $\vartheta_\mathrm{o}$ of the bicone.}
    \label{fig:alldelaymaps_sphericalandbicone}
\end{figure*}

A thin spherical shell may be considered as being composed of thin face-on rings parametrised by $\vartheta \in [0, \pi]$. Without lensing, we see that each thin ring traces out a horizontal line in observed ($v_z, T_X$)-space and the envelope of these horizontal lines is an ellipse. Including lensing, we have that each horizontal line for each thin ring (fixed $\vartheta$) is distorted into an ellipse in ($v_z, T_X$)-space, illustrated in Figure \ref{fig:alldelaymaps_sphericalandbicone}, subplot 5b. These ellipses have the largest height (largest deviation from a straight horizontal line) when $\vartheta = \frac{\pi}{2}$, the middle of the velocity-delay map.

The distortion we describe over the entire velocity-delay map for the thin spherical shell is also very small. For example, we can check how the duration of the time delay of the overall signal is modified. Recognising that the maximum and minimum of the time delay occurs when $\sin (\varphi + \phi_X) =\pm 1$ respectively, and repeating this argument after combining the remaining $\vartheta$-dependent terms into a single sine term, we have
\begin{equation}
        \tfrac{R}{c} (1+z_s)( 1 - K_X )\leq T_X \leq \tfrac{R}{c} (1+z_s)( 1 + K_X ) \label{eq:domainTxsphericalshell}
\end{equation}
where $K_X \equiv \sqrt{1+F_X^2(1+z_s)^{-2}}$ and $F_X^2(1+z_s)^{-2} \sim 10^{-12}$.

If we instead consider radial inflow or outflow (Equation \eqref{radialmotioneqn} with $u(r) <0 $ and $u(r) >0$ respectively), we have Equation \eqref{thinsphericalshelltd} combined with
\begin{equation}
    u_z (\vartheta ) = u(R) \cos \vartheta.
\end{equation}
Since here $u_z$ is a function of $\vartheta$ independent of $\varphi$, it is convenient for interpretation to hold $\varphi$ constant instead of $\vartheta$. Without lensing, the velocity-delay map is an interval. By choosing the observer to be in the positive $z$ direction (such that increasing $v_z$ corresponds with increasing wavelength), spherical outflow and inflow corresponds to a negative and positive slope respectively. When the quasar is lensed, i.e. $F_X >0$, the interval deforms into a filled ellipse, since each fixed value of $\varphi$ corresponds to an arc of an ellipse parametrised by $\vartheta$, the envelope of which is given by $\varphi$ corresponding to $\sin( \varphi + \phi_X) =1$. The largest change in the time delay between lensed images corresponds to where $\vartheta = \frac{\pi}{2}$, i.e. $v_z = 0$, and the magnitude of this change is simply 
\begin{equation}
    \mathcal{T}_B ( \varphi) - \mathcal{T}_A ( \varphi) = \frac{R}{c} \sin \vartheta (F_B \sin ( \varphi + \phi_B) - F_A \sin( \varphi + \phi_A) ).
\end{equation}
However, since $\phi_B \neq \phi_A$, the time delay difference cannot be found from the envelopes of the velocity-delay maps which correspond to different $\varphi$. Again, one can easily build up the expected velocity-delay map for a thick spherical shell or ball model of the BLR, by letting $R$ vary over $R_{\mathrm{min}}$ to $R_{\mathrm{max}}$; this is shown illustrated in Figure \ref{fig:alldelaymaps_sphericalandbicone}.

\subsection{Biconical Geometries}

A biconical BLR geometry is a solid or hollow double cone with a given radial interval $r \in [R_{\mathrm{min}}, R_{\mathrm{max}}]$ and a half-opening angle $\vartheta_\mathrm{o} \in \left(0, \frac{\pi}{2}\right]$; aligned along an axis at an inclination angle $i$ to the line-of-sight direction $\hat{r}_z$. In the hollow case, the BLR material is restricted to $\vartheta = \vartheta_\mathrm{o}, \pi - \vartheta_\mathrm{o}$, and in the solid case it is restricted to $\vartheta \in [0, \vartheta_\mathrm{o}] \cup [\pi - \vartheta_\mathrm{o}, \pi]$. A general biconical geometry is therefore a subset of a thick spherical shell geometry: a solid double cone forms a thick spherical shell when $\vartheta_\mathrm{o} = \frac{\pi}{2}$.

A solid double cone can be constructed from an infinite number of hollow double cones of progressively smaller half-opening angles, and a hollow double cone can be composed from an infinite number of thin rings with radial coordinates between $R_{\mathrm{min}}$ and $R_{\mathrm{max}}$ at $\vartheta = \vartheta_\mathrm{o}$ and $\pi - \vartheta_\mathrm{o}$: the parametric equations for the velocity-delay map for the thin rings at radius $R$ and $\vartheta = \vartheta_\mathrm{o}$ and $\pi - \vartheta_\mathrm{o}$ in radial motion are
\begin{subequations}
\begin{align}
\mathcal{T}_X(\varphi) &= \tfrac{R}{c} \left( 1+z_s \pm J_X \cos \vartheta_\mathrm{o} + F_X \sin \vartheta_\mathrm{o} \sin ( \varphi + \phi_X) \right)\\
u_z(\varphi) &= u(R) ( \pm \cos i \cos \vartheta_\mathrm{o} - \sin i \sin \vartheta_\mathrm{o} \sin \varphi).
\end{align}
\end{subequations}
Without lensing, as $\phi_X \to 0$, each thin ring maps to an interval on the $(v_z, T_X)$-plane; whereas with lensing this interval deforms into an ellipse. The full velocity-delay map from the entire structure is then given by the superposition of such intervals or ellipses; and this velocity-delay map is a subset of the region mapped by a spherical geometry with the same velocity field, dependent on the values of $i$ and $\vartheta_\mathrm{o}$. This is illustrated in Figure \ref{fig:alldelaymaps_sphericalandbicone}, subplots 7c and 7d.

We see that for all of the geometries besides a single thin inclined ring, that the signal from a particular position in the BLR structure is typically superimposed by the signal from many other positions in the structure: the time delay and the line-of-sight velocity alone are insufficient in constraining the BLR position.

\section{Parameter Estimation} \label{sec:parameterestimation}

\begin{figure*}
    \centering
    \renewcommand{\thefigure}{6}
    \includegraphics[width=\linewidth]{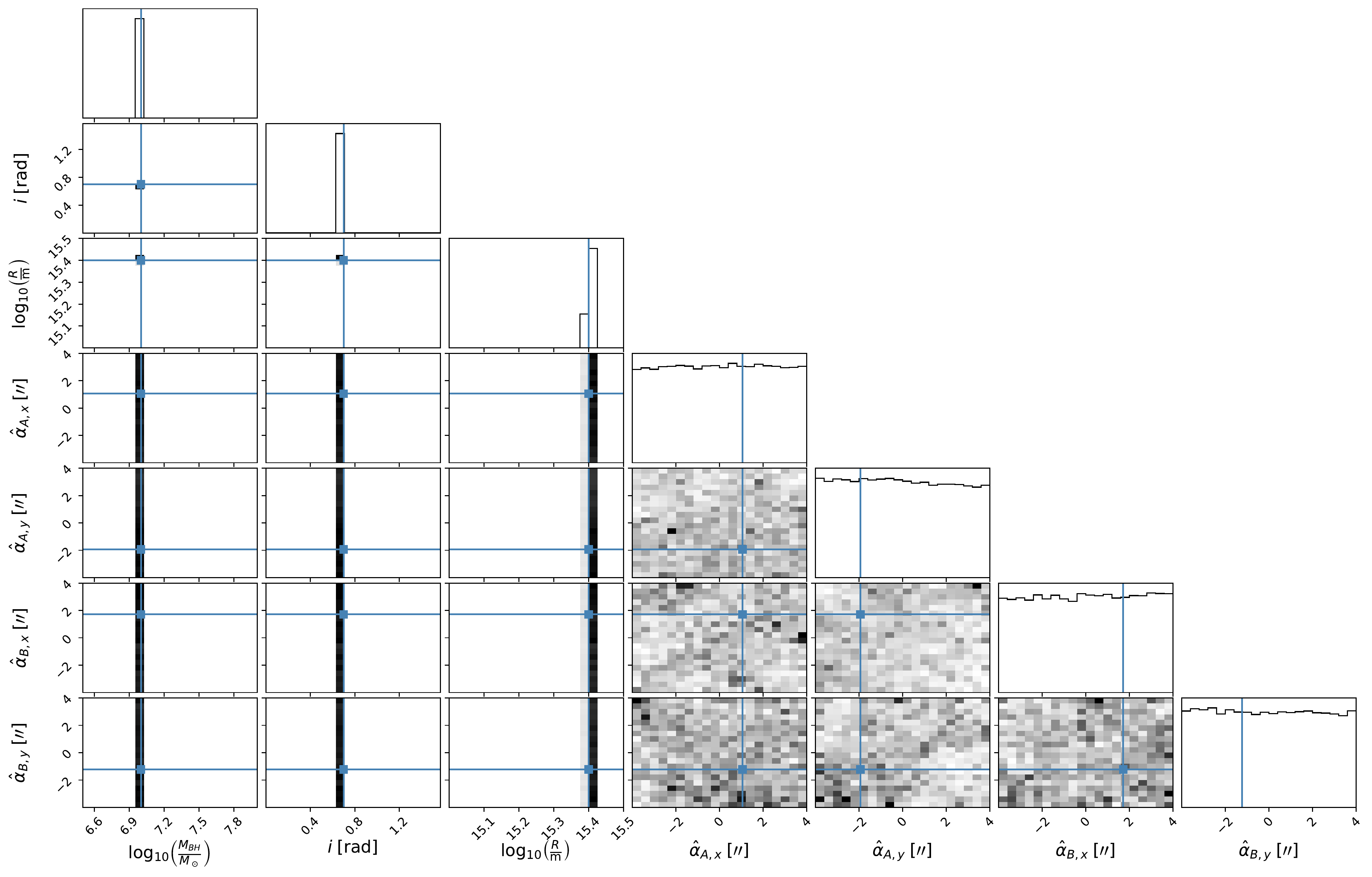}
    \caption{Corner plot showing the one and two dimensional projections of the posterior probability distributions of the parameters $\log_{10}\left(\tfrac{M}{M_{\odot}}\right)$, $i$, $\log_{10}\left(\tfrac{R}{\textrm{m}}\right)$, $\hat{\alpha}_{A,x}$, $\hat{\alpha}_{A,y}$, $\hat{\alpha}_{B,x}$, $\hat{\alpha}_{B,y}$ corresponding to the black hole mass, inclination angle, BLR radius and deflection angles of each image with their true values (7, 0.7 rad, 15.4, 1.05", -1.94", 1.72", -1.23") marked by the blue squares. The range of each subplot corresponds to the bounds of the priors. We assume an idealised scenario where the lensed transfer functions from two images A and B have been recovered (although one is sufficient) and indicate a thin inclined ring in Keplerian orbit. Using the MCMC algorithm {\sffamily emcee} we illustrate that it is possible to recover the desired BLR geometric parameters even if the lensing parameters (i.e. the deflection angles) are not constrained. That the BLR parameters are extremely well constrained is not a comment on reverberation mapping by itself, but attributed to the idealistic model used to illustrate the insensitivity to lensing.}
    \label{fig:cornerplot}
\end{figure*}

Since the available data comes from observations of each unresolved lensed image, the most probable BLR geometric parameters need to be estimated together with lensing parameters. Consequently, the estimates of BLR parameters which are to then be used for estimation of the distance ratio $\frac{D_d}{D_{ds}}$ according to the proposed method are, through the physical deflection angles, implicitly dependent on the source position $\bm{\beta}$. However, we find that it is possible to recover the desired BLR geometric parameters even if the lensing parameters (i.e. the deflection angles) are not constrained, as may be expected from the prior discussions on the smallness of the effect of the lensing for all physically reasonable scales.

As an illustration of this component of the method, we demonstrate finding these parameters using Bayesian parameter estimation, assuming that the transfer function has already been recovered, of the least degenerate, idealistic scenario of contriving the BLR model as restricted to a thin inclined ring. The parameters for this model which we wish to input into the expression for the distance ratio, Equation \eqref{cosmologyeqn_thinring}, are the black hole mass $M$, the inclination angle $i$ and BLR radius $R$.

In Bayesian data analysis, the posterior distribution $\mathcal{P}(\bm{\Theta}_M | \bm{\mathcal{D}}, \mathcal{I})$ is the probability distribution of a given set of parameters $\bm{\Theta}_M$ belonging to a model $M$ given some data $\bm{\mathcal{D}}$ and prior information $\mathcal{I}$. The posterior distribution may be determined through Bayes Rule:
\begin{equation}
    \mathcal{P}(\bm{\Theta}_M | \bm{\mathcal{D}}, \mathcal{I}) = \frac{\mathcal{P}(\bm{\mathcal{D}} | \bm{\Theta}_M , \mathcal{I} ) \mathcal{P} (\bm{\Theta}_M | \mathcal{I})}{\mathcal{P}(\bm{\mathcal{D}}| \mathcal{I})}
\end{equation}
where $\mathcal{P}(\bm{\mathcal{D}}| \bm{\Theta}_M, \mathcal{I})$ is the likelihood, $\mathcal{P}(\bm{\Theta}_M |\mathcal{I})$ is the prior, and the evidence or marginal likelihood is $\mathcal{P}(\bm{\mathcal{D}} | \mathcal{I}) = \int_{\Omega_{\bm{\Theta}_M}} \mathcal{P}(\bm{\mathcal{D}}, \bm{\Theta}_M | \mathcal{I} ) \mathcal{P}(\bm{\Theta}_M | \mathcal{I}) d \bm{\Theta}_M $. The integral is over the domain of the parameter space $\Omega_{\bm{\Theta}_M}$ and as we are only considering a single model, we drop the subscript $M$ for the rest of this work. The posterior may be estimated via numerical methods, usually involving an algorithm that approximates the posterior by generating and weighting a set of discrete samples. Since evaluating the  multi-dimensional integral to compute the evidence is a key difficulty, common numerical methods such as Markov Chain Monte Carlo (MCMC) rely on drawing these samples from a distribution proportional to the posterior distribution. In this work we use the {\sffamily emcee} \citep{Foreman-Mackey2013} implementation of the MCMC algorithm. We assume non-informative priors, such that the posterior is directly proportional to the likelihood
\begin{equation}
    \mathcal{P}(\bm{\Theta} | \bm{\mathcal{D}}, \mathcal{I}) \propto \mathcal{P}(\bm{\mathcal{D}} | \bm{\Theta} , \mathcal{I})
\end{equation}
and the maximum a posteriori estimate of $\bm{\Theta}$ is equivalent to finding the maximum likelihood estimate.

We define our likelihood function, first specifying our model. Let us assume that the original continuum function $C_{\text{original}} (t_X') = \delta( t_X' - t_i)$ is a Dirac Delta function which is then convolved or blurred in the time domain by a Gaussian function such that:
\begin{equation}
    C( t_X') \propto \exp{\left[ - \frac{(t_X' - t_i)^2}{2 \epsilon^2}\right]}
\end{equation}
where $\epsilon$ is the width or standard deviation of the blurred continuum, which allows for numerical sampling of the posterior. Since for a thin inclined ring, the transfer function $\Psi_X (v_z, T_X)$ is given by a sum of Dirac-delta signals  \eqref{spectralfluxthinring}, the line response with $t_i = 0$ becomes
\begin{equation}
\begin{split}
    L_X(v_z, t_X') &\propto \frac{R}{V^+(R, v_z)} \left( \exp{\left[ - \frac{(t_X' -  W_X^+(R, v_z))^2}{2 \epsilon^2}\right]} \right.\\
    &+\left.\exp{\left[ - \frac{(t_X' - W_X^-(R, v_z))^2}{2 \epsilon^2}\right]}\right) \label{modelLX}
\end{split}
\end{equation}
where 
\begin{align}
    V^{+}(r, v_z) &\equiv u'(r) \sin \varphi^+ = \sqrt{u'^2(r) - v_z^2}\\
\begin{split}
    W^{\pm}_X(r, v_z) &\equiv \mathcal{T}_X (r, \tfrac{\pi}{2}, \varphi^\pm)\\
    &= \frac{r}{c} \left( 1+ z_s + A_X \frac{V^{\pm}(r, v_z)}{u'(r)}  + B_X \frac{v_z}{u'(r) }\right)
\end{split}
\end{align}
and
\begin{align}
    u'(r) &\equiv \sqrt{\tfrac{GM}{r}} \sin i \\
    A_X &\equiv (1+z_s) \sin i - (1+z_d) \hat{\alpha}_{X, y} \cos i\\
    B_X &\equiv  -(1+z_d) \hat{\alpha}_{X,x}
\end{align}
and so, fixing $\epsilon$, the parameters associated with this model are $ \bm{\Theta} = \left(\log_{10}\left(\tfrac{M}{M_{\odot}}\right), i, \log_{10}\left(\tfrac{R}{\textrm{m}}\right), \{ \hat{\alpha}_{X,x}, \hat{\alpha}_{X,y} \} \right)$ where $\{ \hat{\alpha}_{X,x}, \hat{\alpha}_{X,y} \}$ denotes the set of all the physical deflection angles over all the available images (through which there is an implicit dependence on $\frac{D_d}{D_{ds}}$ and $\bm{\beta}$). For the purpose of parameter estimation, the model is considered a function of the parameters $\bm{\Theta}$ instead of the variables $(v_z, t_X')$, which are evaluated at values fixed by each data point and then binned: $L_x(v_z, t_X') \to L_{X,jk} (\bm{\Theta})$, where $j$ and $k$ respectively denote the time and line-of-sight velocity bins of the data.

Rather than work with the analytic expression \eqref{modelLX} directly, which due to the ${V^+(R, v_z)}^{-1}$ term is singular when $v_z^2 = u'^2(r)$, we instead obtain $L_{X,jk} (\bm{\Theta})$ numerically, utilising $L_x(v_z, t_X') \propto \frac{\partial^2 N}{\partial u_z \partial \mathcal{T}_X} ( v_z, t_X')$. For each BLR cloud we may draw a position sample from the thin ring number density distribution function. To each sample we then assign a time and a velocity from the time delay and velocity models \eqref{Txthinring} and \eqref{vzthinring}, before performing a 2D Gaussian blur. The resultant time delay and line-of-sight velocity values were binned with bin widths of roughly 1d and 50 km s$^{-1}$ respectively.

We generate mock data by fixing the model parameters at a chosen set of true values $\bm{\Theta}_{\text{true}}$ and adding measurement noise. As photon statistics follow a Poisson distribution, the scaled Poisson distribution corresponding to the flux data has a mean at the model value $L_{X,jk} (\bm{\Theta_{\text{true}}})$ and a variance given by $\sigma_{jk}^2 =  a L_{X,jk} (\bm{\Theta_{\text{true}}}) + \sigma^2_{\textrm{abs}}$, where $a$ is a constant and there is an additional background contribution $\sigma^2_{\textrm{abs}}$ to the noise. The parameters for the noise are chosen to correspond to a peak signal-to-noise ratio of approximately 10 with $a \approx 0.05 \, \text{max}(L_{X,jk} (\bm{\Theta_{\text{true}}}))$. As the photon count is large, the distribution is approximated as Gaussian 
\begin{equation}
    \mathcal{D}_{X,jk} = \mathcal{N}(0, \sigma_{jk}^2) + L_{X,jk} (\bm{\Theta}_{\text{true}})
\end{equation}
and the probability density of an individual datum $\mathcal{D}_{X,jk}$ is
\begin{equation}
    \mathcal{P}(\mathcal{D}_{X,jk} | \bm{\Theta} , \mathcal{I}) = \frac{1}{\sqrt{2 \pi} \sigma_{jk} } \exp{\left[ - \frac{(\mathcal{D}_{X,jk} - L_{X,jk} (\bm{\Theta}))^2}{2 \sigma_{jk}^2} \right]}.
\end{equation}

The likelihood function for the line response from a single image $X$, i.e. the joint probability density function $\mathcal{P}(\bm{\mathcal{D}_X} | \bm{\Theta}, \mathcal{I})$, is given (assuming independent data) by the product of the probability densities of the individual measurements $\mathcal{D}_{X,jk}$:
\begin{equation}
    \mathcal{P}(\bm{\mathcal{D}_X} | \bm{\Theta}, \mathcal{I}) = \prod_{j,k} \mathcal{P}(\mathcal{D}_{X,jk} | \bm{\Theta}, \mathcal{I}).
\end{equation}
The log-likelihood for the line response for a single image $X$ is thus
\begin{align}
    &\ln{\mathcal{P}(\bm{\mathcal{D}_X} | \bm{\Theta}, \mathcal{I} )} = \sum_{j,k} \left( \ln{ \left(\frac{1}{\sqrt{2 \pi} \sigma_{jk}} \right)} - \frac{\left( \mathcal{D}_{X,jk} -  L_{X,jk} (\bm{\Theta}) \right)^2}{2 \sigma_{jk}^2} \right).
\end{align}
To find the posterior taking into account the data from all images (assuming independence), we can either iterate over the data from each subsequent image; or perform the analysis in one step, where the full likelihood function is the product of the individual ones
\begin{equation}
    \mathcal{P} ( \{ \bm{\mathcal{D}}_X \} |  \bm{\Theta} , \mathcal{I} ) = \prod_{X} p( \bm{\mathcal{D}}_X | \bm{\Theta} , \mathcal{I})
\end{equation}
where $\{ \bm{\mathcal{D}}_X \}$ is the set of data from all images.

We ran {\sffamily emcee} until the Markov chain was longer than 100 times the integrated autocorrelation time estimate for the chain for each parameter, with the estimated autocorrelation time changing by less than 1\%. Due to the highly restrictive assumptions of this BLR geometric model, the geometric parameter values are very well determined; whereas the deflection angles are unconstrained with the posterior distributions equivalent to the uniform priors. This result for realistic lensing scales could be surmised from the data by eye; although even if the scale of lensing is increased to unrealistically large magnitudes, the geometric parameter values remain very well determined for this model. The corner plot of the one and two dimensional projections of the posterior probability distributions in the case of galaxy-scale lensing is shown in Figure \ref{fig:cornerplot}.

\section{Conclusions} \label{sec:conclusions}

In summary, \citep{Ng2020} proposed a method of determining $\frac{D_d}{D_{ds}}$ without lens modelling, via reverberation mapping and measuring times delays of differentially lensed, unresolved quasars. This method suffers from the following:

\begin{enumerate}

    \item Reverberation mapping does not uniquely map the location from discrete flares as is required; rather it only gives general constraints on the geometric structure of the BLR. This is a problem of underdetermination.

    \item The projection of the BLR cloud position on the image-image axis is needed, which requires further data (i.e. spatially resolving each image).
    
    \item The time scale for differential lensing is on the order of minutes to days compared to tens to hundreds of days for the reverberation mapping time delay; $ \tau_{\textsc{rm}} \gg | \delta \tau_{AB} |$ for a general BLR geometry.

    \item Finally, the determination of the geometric parameters is not a priori independent of $\bm{\beta}$. However, the effect of differential lensing is sufficiently small ($\tau_{\textsc{rm}} \gg | \delta  \tau_{AB}|$), implying the geometric parameters may be inferred accurately without additional data or lens modelling.
    
\end{enumerate}

Although imaging the BLR structure of an unlensed quasar requires a resolution of $1-10 \mu$ arcseconds, which is still unfeasible with current interferometry, a high angular resolution between the photocentres of redshifted and blueshifted emission may be achieved using the technique of spectroastrometry \citep{Sturm2018}. This has allowed for initial parallax distance measurements using reverberation mapped quasars \citep{Wang2020}. Spectroastrometry would not be as straightforward (and perhaps superfluous) to apply to lensed quasars and obtaining the source positions from the observed image positions would require lens modelling; although the possibility of image magnification could be an advantage. It is possible that differential lensing time delays may be used to extract cosmological information, although obtaining the source positions from image positions may remain a limitation, if distinct, time-variable features within the quasar structure between images can be identified.

\section*{Acknowledgements}
AN thanks Ed McDonald, C\'{e}line B{\oe}hm and Mark Wardle for useful discussions and comments; Mat Varidel and Josh Speagle for discussions on Bayesian analysis; and Oz Brent for discussions regarding numerical computation. AN also thanks the anonymous reviewer for helpful comments which improved this manuscript. This research was funded in part by an Australian Government Research Training Program Scholarship and University of Sydney Hunstead Merit Award.

%%%%%%%%%%%%%%%%%%%%%%%%%%%%%%%%%%%%%%%%%%%%%%%%%%
\section*{Data Availability}
The data underlying this article will be shared on reasonable request to the corresponding author.

%%%%%%%%%%%%%%%%%%%%%%%%%%%%%%%%%%%%%%%%%%%%%%%%%%



\bibliographystyle{mnras}
%\bibliography{time_delay_2} % if your bibtex file is called example.bib
\bibliography{paper2bib}

%%%%%%%%%%%%%%%%%%%%%%%%%%%%%%%%%%%%%%%%%%%%%%%%%%

%%%%%%%%%%%%%%%%% APPENDICES %%%%%%%%%%%%%%%%%%%%%

\appendix

\section{Calculations of Transfer Functions} \label{appendixa}

\subsection{Thin Inclined Ring}
The number density distribution which describes a thin inclined ring BLR of radius $R$ with uniform linear number density $\mu$, an inclination angle $i$ is
$n(\bm{r}) = \mu \delta(r - R)r^{-1} \delta(\vartheta - \frac{\pi}{2})$. 
\subsubsection{One-Dimensional Transfer Function}
The one-dimensional transfer function is proportional to
\begin{align}
    &\frac{\partial N}{\partial \mathcal{T}_X} (T_X) = \iiint n(\bm{r}) \delta (\mathcal{T}_X(\bm{r}) - T_X) \, d^3 \bm{r}\\
    &= \iiint \mu \delta (r - R) \delta (\vartheta - \tfrac{\pi}{2}) \delta ( \mathcal{T}_X(r, \vartheta , \varphi) -T_X)r \sin \vartheta \, dr d\vartheta d \varphi \\
    &= \int \mu R \delta (\mathcal{T}_X(R, \tfrac{\pi}{2}, \varphi) - T_X) d \varphi \\
    &= \int \mu R \sum\limits_{\tilde{\varphi}} \delta (\tilde{\varphi} - \varphi) \left| \frac{\partial \varphi}{\partial \mathcal{T}_X} (R, \tfrac{\pi}{2}, \varphi) \right| \, d\varphi\\
    &= \sum\limits_{\tilde{\varphi}} \frac{\mu c}{\left| F_X \cos ( \tilde{\varphi} + \phi_X) \right|}
\end{align}
where $\tilde{\varphi}$ are the roots of $\mathcal{T}_X(R, \tfrac{\pi}{2}, \varphi)$.
This gives the result
\begin{equation}
    \frac{\partial N}{ \partial \mathcal{T}_X} (T_X) \propto \frac{
    2 \mu c}{\sqrt{F_X^2 - \left(T_X c R^{-1} - (1 +z_s) \right)^2}} \label{fluxthinringAppendix}
\end{equation}
when $\left|T_X c R^{-1} - (1 +z_s) \right| \leq F_X$ and $0$ otherwise.
\subsubsection{Two-Dimensional Transfer Function}
Although computing the two-dimensional transfer function \eqref{eq:spectralflux} for a general $n(\bm{r})$ is not an easy task, the presence of Dirac delta distributions in $n(\bm{r})$ means we can reduce the dimensionality of the integral. The rule for the composition of a Dirac delta distribution function gives a more general version of the usual change of variables rule (permitting the new and old random variables to not be one-to-one). Assuming the ring is in Keplerian orbit with orbital speed $u$, we obtain
\begin{align}
\begin{split}
    &\frac{\partial^2 N}{\partial u_z \partial \mathcal{T}_X} ( v_z, T_X) \\
    &= \mu \! \iint \! \delta(r-R) r \delta^2( u_z(r,\tfrac{\pi}{2}, \varphi) - v_z, \mathcal{T}_X(r,\tfrac{\pi}{2}, \varphi) - T_X)) \, dr d \varphi
    \end{split}
    \\
    &= \mu \! \sum\limits_{\tilde{r}, \tilde{\varphi}} \iint \! \delta (r - R) r \delta^2(r - \tilde{r}, \varphi - \tilde{\varphi}) \left|\frac{\partial(r, \varphi)}{\partial(u_z,\mathcal{T}_X)}(r, \tfrac{\pi}{2}, \varphi) \right| dr d \varphi\\
    &= \mu R \sum\limits_{\tilde{r}, \tilde{\varphi}} \delta (\tilde{r} - R)  \left|\frac{\partial(r, \varphi)}{\partial(u_z,\mathcal{T}_X)}(R, \tfrac{\pi}{2}, \tilde{\varphi}) \right|
\end{align}
where $\tilde{r}$ and $\tilde{\varphi}$ solve $T_X = \mathcal{T}_X(\tilde{r}, \tfrac{\pi}{2}, \tilde{\varphi})$ and $v_z = u_z(\tilde{r}, \tfrac{\pi}{2}, \tilde{\varphi})$. 

The Jacobian determinant $\left|\frac{\partial (r, \varphi)}{\partial (u_z, \mathcal{T}_X)} (r, \tfrac{\pi}{2}, \varphi) \right|= \left|\frac{\partial u_z}{\partial r} \frac{\partial \mathcal{T}_X}{\partial \varphi} - \frac{\partial \mathcal{T}_X}{\partial r} \frac{\partial u_z}{ \partial \varphi} \right|^{-1}$ when evaluated at the two roots $\varphi^\pm \equiv \tilde{\varphi}$ is
\begin{equation}
\begin{split}
    &\left|\frac{\partial (r, \varphi)}{\partial (u_z, \mathcal{T}_X)} (r, \tfrac{\pi}{2}, \varphi^\pm) \right| = \left| \frac{\partial u(r)}{dr} \frac{r}{c} \frac{v_z}{u^2(r) \sin i }  \right. \\
    & \times \left( A_X v_z  - B_X V^\pm (r, v_z) \right) +\left. V^\pm (r, v_z) W_X^\pm (r, v_z) r^{-1} \right|^{-1}
    \end{split} \label{diskdeterminant}
\end{equation}
where we defined
\begin{align}
    V^{\pm}(r, v_z) &\equiv u(r) \sin i \sin \varphi^\pm = \pm \sqrt{(u(r) \sin i)^2 - v_z^2}\\
\begin{split}
    W^{\pm}_X(r, v_z) &\equiv \mathcal{T}_X (r, \tfrac{\pi}{2}, \varphi^\pm)\\
    &= \frac{r}{c} \left( 1+ z_s + A_X \frac{V^{\pm}(r, v_z)}{u(r) \sin i}  + B_X \frac{v_z}{u(r) \sin i }\right).
\end{split}
\end{align}
In the case of a thin ring the radius is fixed: the first term in Equation \eqref{diskdeterminant} vanishes and we find
$\tilde{r} = \frac{R}{W^\pm (R, v_z)} T_X$, giving
\begin{equation}
    \frac{\partial^2 N}{\partial u_z \partial \mathcal{T}_X} ( v_z, T_X) = \mu c R \! \sum\limits_{\pm} \frac{\delta\left(T_X - W^{\pm}_X(R, v_z) \right)}{V^{+}(R, v_z)}
\end{equation}
\subsection{Inclined Disk}
The number density describing a disk of maximum radius $R$ and uniform surface density $\sigma$ can expressed using the Heaviside step function $H$:
\begin{align}
    n(\bm{r}) &= \sigma  H(R - r) r^{-1} \delta (\vartheta - \tfrac{\pi}{2} )\\
    &= \sigma \int_{-\infty}^{0} \delta( r' + R -r ) r^{-1} \delta (\vartheta - \tfrac{\pi}{2})  \, dr'.
\end{align}
\subsubsection{One-Dimensional Transfer Function}
By recalling the expression for the one-dimensional transfer function for a thin inclined ring, Equation \eqref{fluxthinringAppendix}, we recognise
\begin{equation}
 \frac{\partial N}{\partial \mathcal{T}_X} (T_X) \propto 2 \sigma c \int_{r'_{\mathrm{min}}}^{r'_{\mathrm{max}}} \frac{dr'}{\sqrt{F_X^2 - \left(\frac{cT_X}{r' + R} - (1+z_s) \right)^2}} 
\end{equation}
where $r'_{\mathrm{min }} = \mathrm{max} \left[ - \infty, T_X^+  - R\right] = T_X^+ -R$ and $r'_{\mathrm{max}} = \mathrm{min} \left[0, T_X^- -R \right]$ with $T_X^{\pm} \equiv \frac{c T_X}{(1+z_s) \pm F_X}$. Under a change of variables from $r'$ to $y = r' + R$, such that $y_{\mathrm{min }} =  T_X^+$ and $y_{\mathrm{max}} = \mathrm{min} [R,  T_X^-]$, we have that when $F_X < (1+z_s)$,
\begin{align}
& \frac{\partial N}{\partial \mathcal{T}_X} (T_X) \propto \int_{y_{\mathrm{min}}}^{y_{\mathrm{max}}} \frac{2 \sigma c y \; dy}{\sqrt{(1+z_s)^2 - F_X^2} \sqrt{(y - T_X^+)(T_X^- - y)}} \\
\begin{split}
&\propto \frac{2 \sigma c }{\sqrt{(1+z_s)^2 - F_X^2}} \Bigg[- \sqrt{(T_X^- - y)(y- T_X^+)}\\
&\quad \qquad \left. - \frac{(T_X^- + T_X^+)}{2} \arcsin \left(\frac{2y - (T_X^+ + T_X^-)}{(T_X^+ - T_X^-)} \right) \right]^{\mathrm{min}[R, T_X^-]}_{T_X^+}.
\end{split}
\end{align}
When $y_{\mathrm{max}} = T_X^-$, then $T_X < \frac{R}{c}\left( (1+z_s) - F_X \right)$ and the one-dimensional transfer function is a linear function of $T_X$,
\begin{equation}
    \frac{\partial N}{\partial \mathcal{T}_X} (T_X) \propto 2 \pi \sigma (1+z_s) c^2 \left((1+z_s)^2 - F_X^2\right)^{-\tfrac{3}{2}}T_X
\end{equation}
whereas when $y_{\mathrm{max}} = R$, i.e. $\frac{R}{c} \left( (1+z_s) - F_X \right) \leq T_X \leq \frac{R}{c} \left( (1+z_s) + F_X \right)$, the expression does not simplify as nicely as the former regime.

Similarly, in the case that $F_X > (1 + z_s)$,
\begin{align}
    &\frac{\partial N}{\partial \mathcal{T}_X} (T_X) \propto \int_{y_{\mathrm{min}}}^{y_{\mathrm{max}}} \frac{2 \sigma c y \; dy}{\sqrt{F_X^2 - (1+z_s)^2} \sqrt{(y - T_X^+)(y - T_X^-)}}  \\
\begin{split}
    &\propto \frac{2 \sigma c}{\sqrt{F_X^2 -(1+z_s)^2}} \Bigg[ \sqrt{(y - T_X^+)(y - T_X^-)} \\
    &\qquad \qquad + \left. \frac{(T_X^+ + T_X^-)}{2} \mathrm{arccosh} \left( \frac{2y}{T_X^+ + T_X^-} -1 \right) \right]^{ \mathrm{min}[R, T_X^+]}_{T_X^-}
\end{split}
\end{align}
and finally, in the case that $F_X = (1+z_s)$,
\begin{equation}
   \frac{\partial N}{\partial \mathcal{T}_X} (T_X) \propto \frac{2 \sigma \sqrt{-cT_X (c T_X - 2(1+z_s) R)} ((1+z_s)R + cT_X)}{3(1+z_s)^2 T_X}.
\end{equation}

Since the number density for a radially-thick inclined ring can be found by subtracting the number density of a smaller disk from a larger disk, the one-dimensional transfer function for such a geometry is proportional to
\begin{equation}
    \Psi_{X, \text{thick ring}} (T_X) \propto \left. \Psi_{X, \text{disk}} \right|_{R_{\mathrm{max}}} (T_X) - \left. \Psi_{X, \text{disk}} \right|_{R_{\mathrm{min}}} (T_X).
\end{equation}

\subsubsection{Two-Dimensional Transfer Function}
The two-dimensional transfer function is given by
\begin{align}
\begin{split}
    &\frac{\partial^2 N}{\partial u_z \partial \mathcal{T}_X} (v_z, T_X)\\
    &= \iint \sigma H(R -r) r \delta^2 (u_z(r, \tfrac{\pi}{2}, \varphi) - v_z, \mathcal{T}_X(r, \tfrac{\pi}{2}, \varphi) - T_X) \, dr d \varphi
\end{split}\\
    &= \sum\limits_{\tilde{r}, \tilde{\varphi}} \iint \sigma H(R-r) r \delta^2(r - \tilde{r}, \varphi - \tilde{\varphi}) \left| \frac{\partial (r, \varphi)}{\partial (u_z, \mathcal{T}_X)}( \tilde{r}, \tfrac{\pi}{2}, \tilde{\varphi}) \right| \, dr d \varphi.
\end{align}
Using the Jacobian  determinant \eqref{diskdeterminant}, along with $u(r) \propto r^{-1/2}$ we have the result
\begin{equation}
\begin{split}
&\frac{\partial^2 N}{\partial u_z \partial \mathcal{T}_X} (v_z, T_X) = \sum\limits_{\tilde{r}, \pm} c \sigma H(R- \tilde{r}) \tilde{r} \, \bigg| (1+z_s) V^{\pm}(\tilde{r}, v_z)\\
&+  \left( (V^{\pm}(\tilde{r}, v_z))^2 - \tfrac{v_z^2}{2} \right) \frac{A_X}{u \sin i} + \tfrac{3}{2} V^{\pm}(\tilde{r}, v_z) \frac{v_z}{u \sin i} B_X  \bigg|^{-1}.
\end{split}
\end{equation}
As $T_X = \mathcal{T}_X(\tilde{r}, \tfrac{\pi}{2}, \varphi^\pm )$ and $v_z = u_z(\tilde{r}, \tfrac{\pi}{2}, \varphi^\pm)$ together form a polynomial of 6\textsuperscript{th} degree in $\tilde{r}$, there are 6 (real or complex) roots for which there are no algebraic expressions.

\subsection{Thin Spherical Shell}

A thin spherical shell of radius $r=R$ with a uniform surface density $\sigma$ in solid body rotation, which has a number density given by $n(\bm{r}) = \sigma \delta (r - R)$.

\subsubsection{One-Dimensional Transfer Function}
Calculating the one-dimensional transfer function in the case of a thin spherical shell geometry using the direct method is not straightforward due to the nature of the integral. However, the thin spherical shell may be decomposed into thin face-on rings where there is a unique value of $\tau_{\textsc{rm}}$ for each ring. We happen to know that the conditional probability density function of $\delta \tau_X$ for each given $\tau_{\textsc{rm}}$: as we showed in Paper I, the effect of lensing on a thin face-on ring on its flux is to distort each Dirac delta spike into an arcsine distribution whose width is then dependent on $\tau_{\textsc{rm}}$. In this special case, we may integrate over the contributions from each ring at each time $\tau_{\textsc{rm}}$. Formally,  since $\mathcal{T}_X$ is the sum of $\tau_{\textsc{rm}}$ and $\delta \tau_X$ which may be considered as random variables, $\frac{dN}{d\mathcal{T}_X}$ may be written using the definition of the conditional probability as
\begin{align}
    \frac{dN}{d \mathcal{T}_X} (T_X) &= \int\limits_{-\infty}^{\infty} \frac{\partial^2 N}{\partial \tau_{\textsc{rm}} \partial \delta \tau_X} ( T_X - t, t) \, d t\\
    &= \int\limits_{\tau_{\textsc{rm}}^{ \textsc{min}}}^{\tau_{\textsc{rm}}^{\textsc{max}}} f(t) g(T_X - t,  t) \, d t
\end{align}
when $\tau_{\textsc{rm}}^{\textsc{min}}\leq t \leq \tau_{\textsc{rm}}^{\textsc{max}}$ and $\frac{dN}{d \mathcal{T}_X} (T_X) = 0$ otherwise. Here
\begin{equation}
    f(t) \equiv \frac{\partial N}{\partial \tau_{\textsc{rm}}} (t) = \frac{ 2 \pi \sigma R c }{(1+z_s)} 
\end{equation}
when $0 \leq t \leq \frac{2R}{c}(1+z_s)$ and $f(t) =0$ otherwise. The conditional number density at $\delta \tau_X=s$ given $\tau_{\textsc{rm}}=t$ is  $g(s, t)$, 
where
\begin{equation}
    g(s, t) = 
\frac{1 }{ \pi \sqrt{\left(\tfrac{R}{c} F_X \right)^2 \left( 1- t'^2\right) - s^2}} 
\end{equation}
where $t' \equiv 1- \tfrac{c t}{R(1+z_s)}$
when the condition $\left(\tfrac{R}{c} F_X \right)^2 \left( 1- t'^2\right) \geq s^2$ holds and $0$ otherwise. 
This condition on $t$ is equivalent to $\tau_{\textsc{rm}}^-\leq t\leq \tau_{\textsc{rm}}^+,$
where the bounds $\tau_{\textsc{rm}}^{\pm}$ depend on $s.$
%and also gives the domain of $T_X$ in agreement with the expression \eqref{eq:domainTxsphericalshell}.
We therefore find $\tau_{\textsc{rm}}^{\textsc{min}}  = \mathrm{Max} [ 0 , \tau_{\textsc{rm}}^{-}] = \tau_{\textsc{rm}}^{-} $ and $\tau_{\textsc{rm}}^{\textsc{max}} = \mathrm{Min} [ \tfrac{2R}{c}(1+z_s), \tau_{\textsc{rm}}^{+}]$; and write
\begin{align}
    \frac{\partial N}{\partial \mathcal{T}_X} (T_X) &= \frac{2 \sigma cR}{(1+z_s) K_X} \int\limits_{\tau_{\textsc{rm}}^{ \textsc{min}}}^{\tau_{\textsc{rm}}^{\textsc{max}}} \frac{d t}{ \sqrt{(t - \tau_{\textsc{rm}}^{ -})(\tau_{\textsc{rm}}^{+} - t)}}\\
    &= \frac{4 \sigma Rc}{(1+z_s) K_X} \arcsin{\left( \sqrt{\frac{\tau_{\textsc{rm}}^{\textsc{max}} - \tau_{\textsc{rm}}^{-} }{\tau_{\textsc{rm}}^{+} - \tau_{\textsc{rm}}^{-}}} \right)}
\end{align}
recalling $K_X \equiv \sqrt{1+F_X^2(1+z_s)^{-2}}$, which gives the result
\begin{equation}
    \frac{\partial N}{\partial \mathcal{T}_X} (T_X) = \frac{2 \pi \sigma R c}{(1+z_s) K_X} 
\end{equation}
when $\tfrac{R}{c}(1+z_s)\left( 1- K_X \right) \leq T_X \leq \tfrac{2R}{c}(1+z_s)$ and
\begin{equation}
    \frac{\partial N}{\partial \mathcal{T}_X} (T_X) = \frac{4 \sigma Rc}{(1+z_s) K_X} \arcsin{\left( \sqrt{\frac{\frac{2R}{c}(1+z_s) -\tau_{\textsc{rm}}^{-}(T_X)}{\tau_{\textsc{rm}}^{+}(T_X) - \tau_{\textsc{rm}}^{-}(T_X)}}\right)}
\end{equation}
when $\tfrac{2R}{c} (1+z_s) \leq T_X \leq \tfrac{R}{c} (1+z_s) \left( 1+ K_X \right)$, and we emphasise the dependence of $\tau_{\textsc{rm}}^{\pm}$ on $T_X$. This gives the domain of $T_X$ in agreement with the expression \eqref{eq:domainTxsphericalshell}. We again note that since $K_X \sim 1$, the lensing effect is extremely small for all images.

\subsubsection{Two-Dimensional Transfer Function}

The two-dimensional transfer function is proportional to
\begin{align}
\begin{split}
    &\frac{\partial ^2 N}{\partial u_z \partial \mathcal{T}_X} (v_z, T_X)\\
    &= \sigma R^2 \iint  \delta^2 (u_z(R, \vartheta, \varphi) - v_z, \mathcal{T}_X(R, \vartheta, \varphi) - T_X) \sin \vartheta  \, d \vartheta d \varphi
\end{split}
    \\
    &= \sigma R^2 \sum\limits_{\tilde{\vartheta}, \tilde{\varphi}}\iint \delta^2(\vartheta - \tilde{\vartheta}, \varphi - \tilde{\varphi}) \left| \frac{\partial (\vartheta, \varphi)}{\partial (u_z , \mathcal{T}_X)} (R, \tilde{\vartheta}, \tilde{\varphi}) \right| \sin \vartheta  \, d \vartheta d \varphi\\
    &= \sigma R^2  \sum\limits_{\tilde{\vartheta}, \tilde{\varphi}} \left| \frac{\partial (\vartheta, \varphi)}{\partial (u_z, \mathcal{T}_X)} (R, \tilde{\vartheta}, \tilde{\varphi}) \right| \sin \tilde{\vartheta}\\
    &= \sum\limits_{\tilde{\vartheta}, \tilde{\varphi}} \frac{\sigma Rc}{\left|u \sin i_v \left( F_X \cos \tilde{\vartheta} \cos \phi_X + (1+z_s) \sin \tilde{\vartheta} \sin \tilde{\varphi} \right) \right|}\\
    &= \sum\limits_{\tilde{\varphi}} \frac{\sigma Rc}{\left|F_X \sqrt{(u \sin i_v)^2 - \frac{u^2}{\cos^2  \tilde{\varphi}}} \cos \phi_X  + (1+z_s) u \tan \tilde{\varphi} \right|}
\end{align}
where $\tilde{\vartheta}$ and $\tilde{\varphi}$ solve $T_X = \mathcal{T}_X(R, \tilde{\vartheta}, \tilde{\varphi})$ and $v_z = u_z(R,\tilde{\vartheta}, \tilde{\varphi})$. Finding the roots requires solving \begin{equation}
\begin{split}
T_X' = &- \frac{F_X v_z'}{(1+z_s)} \left( \sqrt{ \cos^{-2} \tilde{\varphi}- \cos^2 \tilde{\varphi}} \cos \phi_X + \sin \phi_X \right)\\
&\pm \sqrt{1 - \frac{v_z'^2}{\cos^2 \tilde{\varphi}}},
\end{split}
\end{equation} 
where $T_X' \equiv 1 - \frac{c T_X}{R(1+z_s)}$ and $v_z' \equiv \frac{v_z}{u \sin i_v}$, for $\tilde{\varphi}$ in terms of $v_z$ and $T_X$.

Solving for $\tilde{\varphi}$ in terms of $v_z$ and $T_X$ is not straightforward. However, since $\frac{F_X}{(1+z_s)}$ and $\phi_X^+$ are very small compared to $T_X'$ we can take a first order approximation of the solution to $\tilde{\varphi}$ such that $\cos^2 \tilde{\varphi} \approx v_z'^2 (1-T_X'^2)^{-1}$, giving
\begin{equation}
\Psi_X (v_z, T_X) \appropto \frac{2 \sigma Rc |v \sin i_v|^{-1} }{\left| F_X T_X' \cos \phi_X + (1+z_s) \sqrt{1 - T_X'^2 - v_z'^2} \right|}.
\end{equation}
When $F_X =(1+z_d)\hat{\alpha}_X =0$, this is the exact expression for the unlensed transfer function, and is an arcsine distribution in both $v_z$ and $T_X$. From this we see that the effect of an, e.g., larger value for $F_X$ is to depress the approximately arcsine distribution; this effect is extremely small around the centre of the distribution but removes the singularity in the distribution at $v_z' = \pm  \sqrt{1-T_X'^2}$.

%%%%%%%%%%%%%%%%%%%%%%%%%%%%%%%%%%%%%%%%%%%%%%%%%%

% Don't change these lines
\bsp	% typesetting comment
\label{lastpage}
\end{document}